\documentclass[manuscript]{aastex}

\newcommand{\reva}{\null}
\newcommand{\finreva}{\null}

\newcommand{\Ts}{$T_{\rm s}$}
\newcommand{\Tex}{$T_{\rm ex}$}

\newcommand{\kms}{km~s$^{-1}$}
\newcommand{\Hmol}{H$_{2}$}

\newcommand{\junit}{ergs cm$^{-2}$ s$^{-1}$ Hz$^{-1}$ sr$^{-1}$}
\newcommand{\lya}{Ly$\alpha$}
\newcommand{\nh}{$N$(H I)}

\newcommand{\spose}[1]{\hbox to 0pt{#1\hss}}
\newcommand{\simlt}{\mathrel{\spose{\lower 3pt\hbox{$\mathchar"218$}}
     \raise 2.0pt\hbox{$\mathchar"13C$}}}
\newcommand{\simgt}{\mathrel{\spose{\lower 3pt\hbox{$\mathchar"218$}}
     \raise 2.0pt\hbox{$\mathchar"13E$}}}
\renewcommand{\ion}[2]{\ifmmode 
\mbox{\rm #1\,{\small\uppercase\expandafter{\romannumeral #2}}}
\else\mbox{#1\,{\small\uppercase\expandafter{\romannumeral #2}}}\fi}

\newlength{\defbaseskip}
\shorttitle{Neutral gas at $z=1.7764$}
\shortauthors{Carswell et al.}

\begin{document}

\title{A cold component and the complex velocity structure of DLA$1331+170$}

\author{R. F. Carswell, R. A. Jorgenson}
\affil{Institute of Astronomy, Madingley Road, Cambridge CB3 0HA, UK}
\email{rfc@ast.cam.ac.uk}

\author{A. M. Wolfe}
\affil{Center for Astrophysics \& Space Sciences, University of California San Diego, 9500 Gilman Drive, La Jolla, CA 92093-0424}

\and 

\author{M. T. Murphy}
\affil{Centre for Astrophysics and Supercomputing, Swinburne University of Technology, Melbourne, Victoria 3122, Australia}



\begin{abstract}

{\reva We examine the velocity structure in the gas associated with \ion{H}{1}
in the
damped Ly$\alpha$ absorption system at redshift $z=1.7764$ towards the
QSO $1331+170$ using Arecibo \ion{H}{1}\,21cm data, optical spectra from
Keck HIRES and ESO VLT-UVES, and a previously published HST STIS ultraviolet
spectrum. From the optical data we find at least two, and possibly three, 
components showing \ion{C}{1} lines. One of these has very narrow lines with 
Doppler parameter $b=0.55${\kms}, corresponding to a kinetic temperature of 
220K if the broadening is thermal and with a 2-$\sigma$ upper limit of 480K. 
We re-examine the H$_2$ analysis undertaken
by \citet{Cui05} using the neutral carbon velocity structure, and find a model
which is, unlike theirs, consistent with a mixture of collisional and 
background radiation excitation of the observed H$_2$ rotational levels.

Using Voigt profile fits to absorption lines from a range of singly ionized 
heavy elements we find eight components covering a velocity range of 
$\sim 110$ {\kms}, with a further outlier over 120 {\kms} away from the 
nearest in the main group. The \ion{H}{1} structure is expected to 
follow some combination of the
singly ionized and neutral gas, but the 21cm absorption profile is 
considerably different. We suggest, as have others, that this may be 
because the different extent and brightness distributions of the radio 
and optical background sources mean that the
sightlines are not the same, and so the spin temperature derived by 
comparing the Ly$\alpha$ and 21cm line strengths has little physical meaning.
The neutral and singly ionized heavy element line profiles also show significant differences,
and so the dominant components in each appear to be physically distinct. Attempts to use the range of atomic masses to separate thermal and turbulent components of their Doppler widths were not generally successful, since there are several blended components and the useful mass range (about a factor of two) is not very large.

The velocity structure in
all ionization stages up to $+3$, apart from the neutral heavy elements, is sufficiently
complex that it is difficult to separate out the corresponding velocity components for
different ionization levels and determine their column densities. 
\finreva}

\end{abstract}


\section{Introduction}

It has long been known that there is significant temperature structure
in the interstellar medium in the Galaxy. Theoretical models, as
originally proposed by e.g. \citet{Fie69} and \citet{McK77} and
refined by several authors subsequently, predict a number of phases in
pressure equilibrium.  These are a cold neutral medium (CNM),
characterized by temperatures $T$ of up to a few 100~K; warm neutral
and warm ionized media (WNM and WIM) with $1000\simlt T\simlt
10000$~K, and a hot ionized medium (HIM) with $T\simgt 10^6$K.

Within the Galaxy, observations of the 21 cm line have led to the
detection of hyperfine spin temperatures $\sim 3000$ - 8000~K,
characteristic of the warm neutral medium (Carilli, Dwarakanath \&
Goss, 1998, Kanekar et al., 2003). Observations of both the CNM and
WNM are described by \citet{Hei03}, who give a range of spin
temperatures for the cold medium up to $\sim 200$~K, and upper limit
kinetic temperatures for the WNM of up to $\sim 20~000$~K. Contrary to
the predictions of \citet{McK77} a significant fraction of the WNM is
at temperatures where the gas is thermally unstable; i.e. $T$
$\approx$ 500 to 1500 K .  On the other hand, direct observational
confirmation of the temperatures predicted for a thermally stable WNM
have been provided by \citet{Red04}. They used high resolution spectra
from the Hubble Space Telescope and compared line profiles from
\ion{D}{1} with those from heavy elements including \ion{Mg}{2} and
\ion{Fe}{2} to obtain local interstellar medium temperatures $\sim
6500\pm 1500$~K. They also noted that the line broadening is not
always purely thermal, and that there is a mean bulk flow component
within regions of a little over $\sim 2$~km~s$^{-1}$.

For most heavy element quasar absorption systems the gas is ionized
and the kinetic temperature estimates range from $\sim 10^4$\,K and up
to $\sim 10^6$\,K in some cases where \ion{O}{6} is found.  For those
with high \ion{H}{1} column densities, the damped Ly$\alpha$ systems
(DLA), the gas is clearly neutral, with the heavy elements neutral or
singly ionized. However, the ubiquitous presence of \ion{C}{4} and
\ion{Si}{4} \citep{Wol00} and the recent detection of \ion{O}{6} and
\ion{N}{5} \citep{Fox07} absorption indicates that these high redshift
gas layers are also multi-phase media. The \ion{H}{1} 21\,cm spin
temperatures provide a means of estimating the thermal
conditions in the gas, and generally spin temperatures are measured to
be $T_{\rm s}\simgt 10^3$~K (see e.g. the compilation by Curran et
al., 2005). On the other hand these estimates assume that the
\ion{H}{1} column density, $N$(\ion{H}{1}), inferred from the
Ly$\alpha$ absorption can be used to compute $T_{\rm s}$ from the
21\,cm optical depth. While this is a straightforward procedure which
has been used for the past 30 years (see Wolfe \& Davis, 1979), it is
complicated by the disparity between the physical sizes of the optical
and radio beams subtended by the background quasar at the absorption
redshift. Whereas the optical beams are typically less than one light
year across, VLBI experiments reveal that the radio beams are
typically a few hundred pc at the low frequencies of the radio
absorption lines {\reva (e.g. Briggs et al., 1989; Polatidis et al.,1995; 
Thakkar et al., 1995). \finreva} Moreover, 
the appearance of shifts in interferometer phase
between the 21\,cm line and continuum in VLBI experiments demonstrates
a shift in the brightness centroid of the background radio source,
which can only be reproduced by a non-uniform distribution of 21\,cm
opacity toward radio sources with scale-lengths of over 200\,pc
(e.g. Wolfe et al., 1976). For these reasons, the
values of $T_{\rm s}$ inferred by these techniques should be regarded
as conservative upper limits.

Temperatures may also be inferred from excitation conditions in other
ions and from molecular hydrogen. For example, using measurements of
the fine structure levels of \ion{Si}{2} and \ion{C}{2} \citet{How05}
found that the excitation conditions constrain the temperature to
$T<954$\,K in one high redshift DLA.  In those DLAs with H$_{2}$
absorption, analyses of the relative populations of the lower
rotational states give kinetic temperatures, which are generally less
than $\sim 200$~K (e.g Cui et al., 2005; Ledoux, Petitjean \&
Srianand, 2006).

The relationship between excitation temperatures $T_{\rm ex}$ and
kinetic temperatures, $T$, depends on the balance between collisional
and radiative processes. In DLAs exhibiting 21 cm absorption the
hyperfine excitation temperature, i.e.  the spin temperature {\Ts},
likely equals the gas kinetic $T$ because the hyperfine levels are
populated mainly by collisions owing to the moderately high densities
for the gas, and the long radiative lifetime of the upper ($F$=1)
hyperfine state. 
Furthermore, even in regions with low density,
ambient {\lya} radiation can couple {\Ts} to $T$ via the
Field-Wouthuysen effect (e.g. Field, 1959). Within the Galaxy, there
is good agreement between $T$ and {\Tex} determined from the relative
populations of the $J$=0 and $J$=1 rotational states of H$_{2}$, an
agreement that holds for column densities $\log N({\rm H}_2)\simgt 16$
(cm$^{-2}$). However, excitation temperatures deduced from the
relative populations of levels with $J \ {\ge}$ 2 depart from $T$
since spontaneous photon emission rates exceed collisional
de-excitation rates due to the short radiative lifetimes. Rather the
values of {\Tex} inferred for these states is determined by the
ambient radiation fields and other processes responsible such as
H$_{2}$ formation on grain surfaces.

In this paper {\reva we obtain upper \finreva}
limits to the gas kinetic temperature in neutral regions using the
\ion{C}{1} absorption lines arising in a DLA at redshift $z=1.77642$
towards the quasar Q$1331+170$. Because this DLA has been detected in
21\,cm absorption \citep{Wol79}, we can compare this temperature with
the spin temperature of $T_{\rm s}\simgt 1000$\,K. Also the H$_2$
excitation temperature {\Tex}$\equiv T_{{\rm H}_2}=150$~K
\citep{Cui05} has been obtained for this DLA.  We shall argue that the
true kinetic temperature of some of the gas is at most a few 100\,K,
and that the value of the spin temperature is likely to be misleading
as a result of non-uniform coverage of the background radio source.
We also re-examine the physical processes responsible for thermal
equilibrium in this DLA and remark on a dilemma not previously
discussed.  Namely, for a plausible range of volume densities,
background radiation alone \citep{Haa01} will drive {\Tex} $>>$ $T$.
For this reason, the reported H$_{2}$ level populations in this DLA
are difficult to understand. However, adopting the velocity structure
inferred from the \ion{C}{1} and applying it to the H$_2$ does allow
us to produce a physically consistent picture.

{\reva We also examine the singly ionized heavy elements, and two more highly 
ionized species, to look for corresponding structure. This has been done for
several objects as part of large DLA surveys e.g. by \citet{Pro99} and 
\citet{Wol00}, with the general conclusion that the singly ionized species 
structure
is similar to that of \ion{Al}{3}, while the component structure of e.g. 
\ion{C}{4} and singly ionized species bear little relation to each other.
We find generally similar results, and show that, for DLA$1331+170$, the
fitting of complex velocity structures and associating components across
ionization levels is not always straightforward. Consequently there
will be uncertainties in the ionization 
equilibrium models of DLAs \citep{How99}. Also, the \ion{Al}{2}/\ion{Al}{3}
indicator has been used for the analysis of sub-DLA
systems (e.g. Dessauges-Zavadsky et al., 2003). 
\finreva}

\section{Observations}

\subsection{Radio}

The hitherto unpublished radio spectrum was acquired with the Arecibo
300\,m telescope of the National Astronomy and Ionosphere Center near
Arecibo, Puerto Rico by one of us (AMW) during April, 1991. Because of
improvements in receiver technology, the signal-to-noise ratios
(S/N) of the data were significantly improved over the Arecibo
discovery spectrum of \citet{Wol79}. The S/N for the data varies
between 200 and 400 per 0.7 km\,s$^{-1}$ pixel, and the resolution after
Hanning smoothing is about 1.4 km\,s$^{-1}$.

\subsection{Optical}

The Keck HIRES \citep{Vog94} spectrum of Q$1331+170$ {\reva is the one
described by \citet{Pro97}. It covers \finreva} the spectral range from 4220 -
6640\,{\AA} at a resolution (full width half maximum) of
6.25~km~s$^{-1}$, with a number of gaps at wavelengths $>
5515$\,{\AA}. The S/N$\sim 100$ per 2~km~s$^{-1}$ pixel at
5400\,{\AA}, decreasing to $\sim 60$ in the 4300 - 4600\,{\AA} region.

The VLT UVES \citep{Dek00} spectrum is based on data available in the
ESO archive in July, 2006. {\reva This was obtained with two of the
standard echelle settings: on 2002-10-04, for ESO program
67.A-0022(A), 3$\times 4500$s exposures with blue arm central
wavelength 346nm and red arm central wavelength 860nm, and on
2003-03-12, for ESO program 68.A-0170(A), $2\times 3600$s eposures
with the blue 390/red 564 setting. $2x2$ on-chip binning was used in both
cases. See the VLT UVES
handbook\footnote{http://www.eso.org/sci/facilities/paranal/instruments/uves/doc/index.html}
for more details. The data were extracted and combined using the UVES\_popler
package\footnote{http://astronomy.swin.edu.au/$\sim$mmurphy/UVES\_popler.html}. 
The combined spectrum \finreva} covers the range
from 3050\,{\AA} to 10080\,{\AA}, with gaps at 4517 - 4623, 5597 -
5679, and some more at wavelengths $> 6650$\,{\AA}. The resolution is
7~km~s$^{-1}$, and the S/N$\sim 40$ per 2.5~km~s$^{-1}$ pixel at
5400\,{\AA}.  At 4300\,{\AA} S/N$\sim 30$, but it is better at shorter
wavelengths, $\sim 35$ at 3600\,{\AA}.

Voigt profiles convolved with the instrument profile were fitted to
the data using the VPFIT package\footnote{http://www.ast.cam.ac.uk/$\sim$rfc/vpfit.html}, version 9.5.
In versions 9 and later the model profiles are generated on a finer
grid than the data, and in the application here at least 9 sample
points were required across the intrinsic FWHM of the narrowest line so
that the intrinsic line profile was adequately modelled. The
convolution with the instrumental profile is done before re-sampling
back to the original pixel scale for direct comparison with the
original data, so there should be no numerical problems associated with
possible undersampling of the absorption profile which might have
arisen with previous versions {\reva of the program. \finreva}

\section{HI spin temperature}

\begin{figure}
\includegraphics[scale=0.67,angle=270]{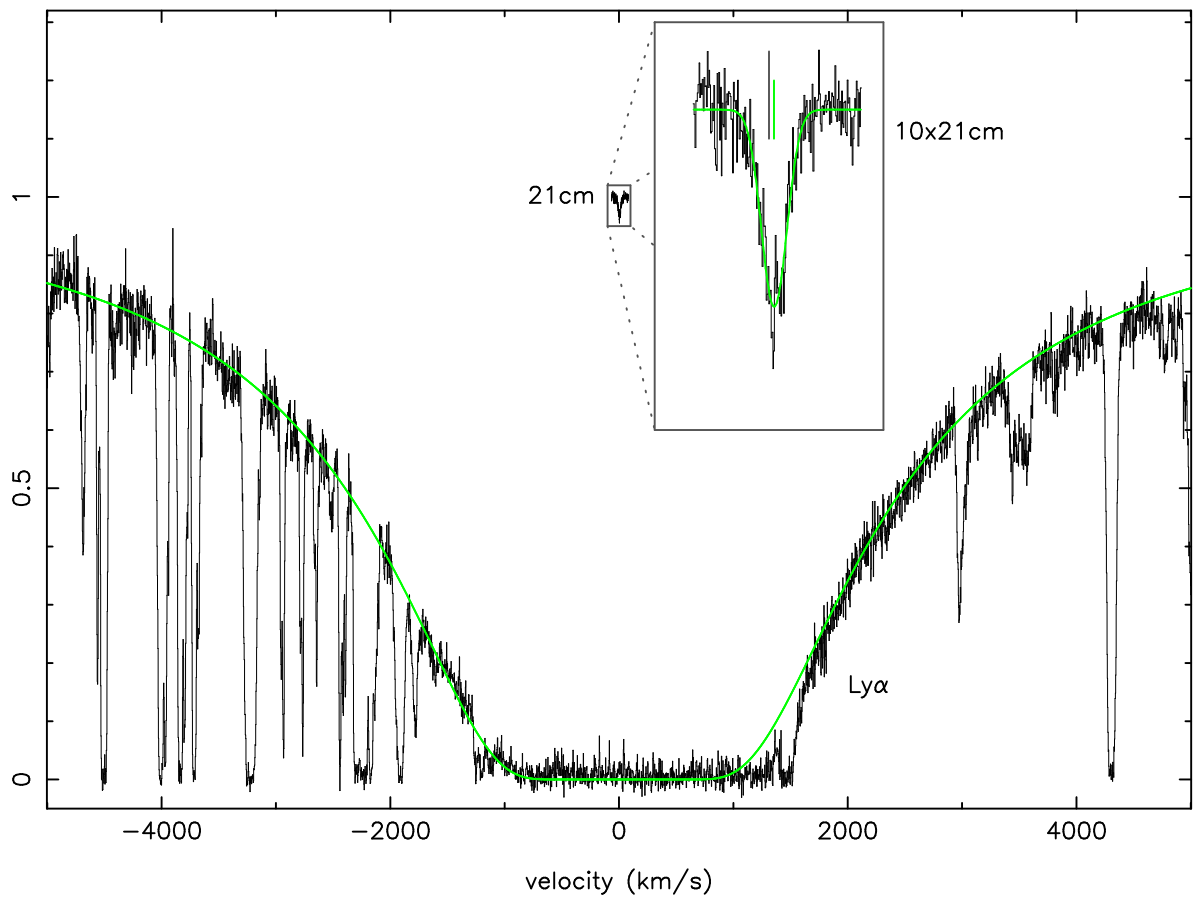}
\caption{\footnotesize The \ion{H}{1} damped Ly$\alpha$ and 21\,cm absorption
line profiles shown against unit continuum on the
same velocity scale relative to a reference redshift $z=1.77642$. The
data are shown in black with Voigt profile fit to the damped Ly$\alpha$ at 
$z=1.77674$
shown in green. The fitted curve departs from the data at longer wavelengths 
in the base of the damped Ly$\alpha$ line because of the presence of another
(sub-DLA) absorption system at $z=1.78636$. The 21\,cm line is also shown 
offset to the right and expanded by a factor of 10 in both $x$- and $y$- 
directions, with a grey tick mark indicating the reference redshift and a 
green one the fitted line centroid. 
\label{fig:HI}}
\end{figure}

While there appears to be some structure in the 21\,cm line profile
(see Figs \ref{fig:HI} and \ref{fig:ccpts}), it is at a marginal level. The overall profile is
well fitted by a single Gaussian with $z=1.7764610\pm 0.0000039$,
$b=15.7\pm 0.7$\,km~s$^{-1}$. Integrating over the optical depth of
the entire 21\,cm feature gives $\int\tau_\nu
dV=0.93\pm 0.03$~km~s$^{-1}$.

The \ion{H}{1} column density inferred from a Voigt profile fit to the
damped Ly$\alpha$ line for the whole complex fitted as a single
component (with an additional component in the long wavelength base to
allow for the known sub-DLA system at $z=1.78636$) is $\log
N$(\ion{H}{1})$=21.17$ (cm$^{-2}$), in agreement with \citet{Pro99}
who report $\log N$(\ion{H}{1})$=21.176\pm 0.041$. The Ly$\alpha$ and 21\,cm 
profiles are shown on the same velocity scale in Fig \ref{fig:HI}.
The formal error in
our coloumn density value is $<0.01$, but systematic errors are likely to
predominate. Estimates using different continua and blending from
other systems gave a range $21.10 < \log N$(\ion{H}{1})$< 21.24$, so
we adopt $0.07$ as an error estimate. The redshift is $z=1.77674\pm
0.00009$, which is consistent with the 21\,cm redshift. If we
constrain the Ly$\alpha$ redshift and Doppler parameter to be the same
as those for the 21\,cm line, then $\log N$(\ion{H}{1})$=21.17$.

With this value for the \ion{H}{1} column density, the spin
temperature $T_{\rm s}=870\pm 160$~K, assuming complete coverage of the
absorber.  This is consistent with the results of \citet{Wol79}, who
give $T_{\rm s}=980$\,K, with a lower limit of 770\,K. If we take the 
covering factor estimate $f=0.72$ provided by \citet{Kan09}, then the
corresponding temperature is $T_{\rm s}=1360$\,K.

\section{Neutral Heavy Elements}
\label{sectCI}

We first investigate transitions arising from the neutral
state of various elements, since these are expected to arise
in the CNM phase of the gas. The main lines available in 
the spectral range covered are from \ion{C}{1} and \ion{Mg}{1}.

\ion{C}{1} in this system has been studied by \citet{Son94}, using
Keck HIRES spectra of the ground state and fine structure lines in the
multiplets at 1656 and 1560\,{\AA} to determine excitation
temperatures of $T_{\rm ex}=10.4\pm 0.5$ for the component at
$z=1.77638$, and $T_{\rm ex}=7.4\pm 0.8$ at $z=1.77654$. The UVES
spectra extend the coverage so that lines at shorter wavelengths,
specifically 1328, 1280 and 1277\,{\AA}, may also be used to constrain
the component parameters. Some of these are blended with lines of
other elements at different redshifts, and all lie within the
Ly$\alpha$ forest absorption region. We have carefully fit the
\ion{C}{1} lines and included possible blends in the fitting
procedure, taking advantage of other transitions from these blended
lines to help constrain their parameters. Details for each wavelength
region used in the fitting procedure are as follows:

\begin{itemize}

\item \ion{C}{1}~1656 falls in a gap in the VLT UVES coverage, and so
  the fit relies on Keck HIRES data alone. There is blending with weak
  \ion{C}{4}~1550 at $z=1.96616$ in the blue wing.  \ion{C}{4}~1548 at
  this redshift shows a single component only, so the 1550 line has no
  significant effect on \ion{C}{1}~1656 at $z=1.77637$.

\item \ion{C}{1}~1560 is covered by both the Keck and VLT spectra. It
  has \ion{Al}{3}~1862 nearby, at $z=1.32535$-$1.32884$, and the
  corresponding \ion{Al}{3}~1854 is blended with CIV 1548 at
  $z=1.78586$-$1.78719$.  The corresponding \ion{C}{4}~1550 shows that
  \ion{Al}{3} blending of \ion{C}{1}~1560 is not
  significant. \ion{C}{1}*~1560 is just shortward of \ion{Fe}{1}~2484
  at $z=0.74461$, from a well-known system which shows multi-component
  \ion{Mg}{2}. \ion{Fe}{1}~2523 \& 2719 were fitted simultaneously at
  this redshift, showing a single velocity component which does not
  affect the \ion{C}{1}* lines.

\item \ion{C}{1}~1328 and lines at shorter wavelength have coverage
  only from UVES at the VLT. The fit to \ion{C}{1}~1328 includes a
  broad weak Ly$\alpha$ at $z=2.03594$, but this changes the effective
  continuum for the \ion{C}{1}* lines mainly. It is much broader than
  the \ion{C}{1}* lines measured.  \ion{C}{1}**~1329 is in the wing of
  Ly$\alpha$ at $z=2.03696$, again with small effect on the line
  parameters.

\item \ion{C}{1} and \ion{C}{1}*~1280 are blended with strong
  Ly$\alpha$ lines at $z=1.92398$ -1.92471. Only the ground state
  \ion{C}{1}~1280 at $z=1.77637$ is well constrained.

\item \ion{C}{1}~1277 is blended with weak Ly$\alpha$ at $z=1.91745$,
  1.91782 and 1.91837, affecting mainly the \ion{C}{1}* and
  \ion{C}{1}** lines with redshifts $z>1.7765$ and rest wavelengths $>
  1277.5$~A.

\end{itemize}

A further complication is that the isotope shifts for
$^{13}$\ion{C}{1} relative to $^{12}$\ion{C}{1} lines at 1656, 1560
and 1328\,A are, respectively, 0.65, -3.30 and
-1.65~km\,s$^{-1}$. These will be particularly important where the
lines are intrinsically narrow. For this reason $^{13}$\ion{C}{1} was
included as a separate species, with wavelengths and oscillator
strengths, for those three lines from \citet{Mor03}. For the other two
transitions used to fit the \ion{C}{1}, at 1280 and 1277~A, the
spectral S/N is lower so the absence of a $^{13}$\ion{C}{1} component
makes little difference to the final result. We verified this by
noting that the removal of $^{13}$\ion{C}{1} 1328 (where the S/N is
higher than for the 1277 or 1280 lines) from the line list had an
insignificant effect on the results.

\begin{figure}
\epsscale{.66}
\plotone{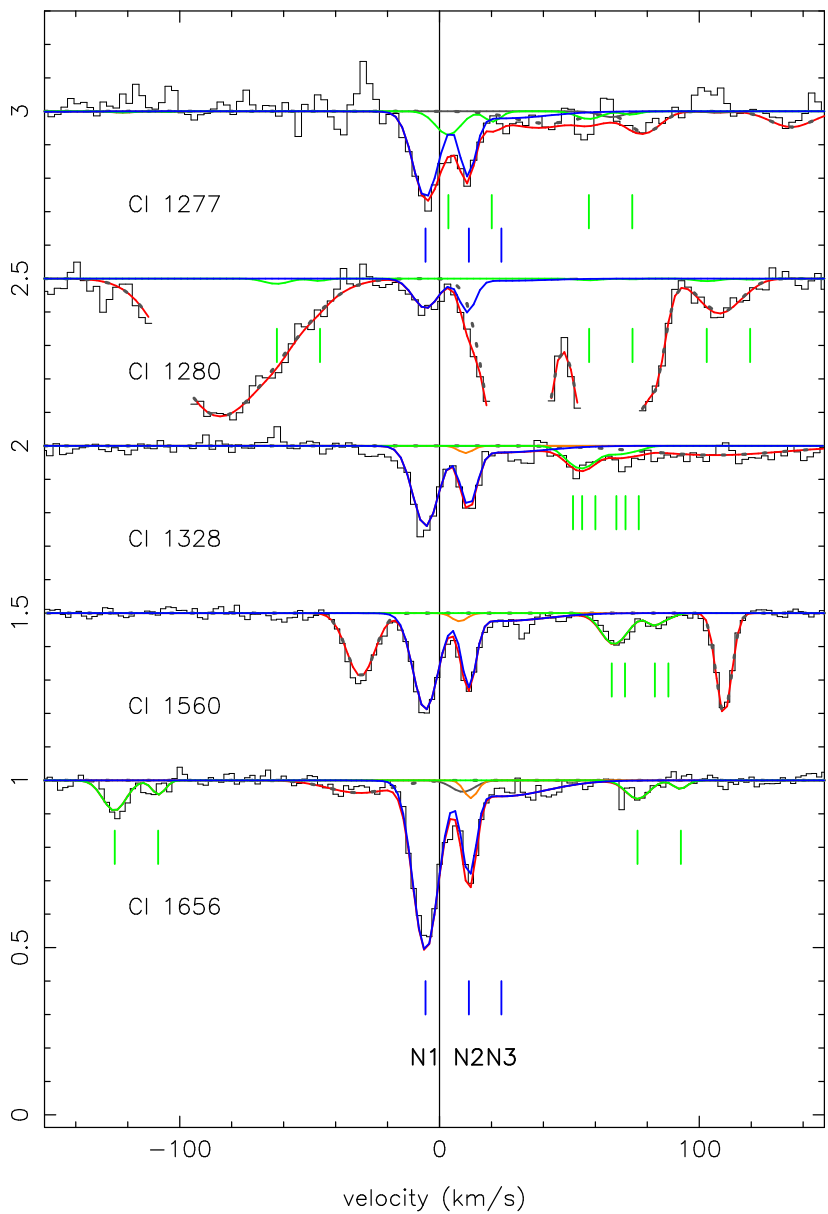}
\caption{\footnotesize The \ion{C}{1} lines 
on a  velocity scale relative to a reference redshift $z=1.77642$. The
data are shown in black with Voigt profile fits to \ion{C}{1} in dark blue, 
\ion{C}{1}* in green,
\ion{C}{1}** as dark blue dot-dash (discernible only for the N1 component 
in the
\ion{C}{1}~1656 profile at $\sim 5$~\kms), and unrelated lines at
different redshifts as grey dots. The possible $^{13}$\ion{C}{1} component 
is shown in orange. The total fitted profiles are shown in red. The continuum is normalized to unity, and different
zero offsets (of an integer $\times 0.5$) have been applied to
separate the various transitions. The tick marks show the centroids of
the fitted features for the
\ion{C}{1} components N1, N2 and N3 (see text). The green tick marks 
indicate the positions for the \ion{C}{1}$^*$ transitions in each wavelength 
region.
\label{fig:CI}}
\end{figure}

The results of the profile fitting are shown in Fig \ref{fig:CI}. A
satisfactory fit to the data required the presence of at least three
\ion{C}{1} velocity components, at velocities of $-5.4$, 11.3 and 23.8
km~s$^{-1}$ relative to an arbitrarily chosen reference redshift of
$z=1.77642$.  We henceforth designate these as components N1, N2, and N3
respectively.  Details of the fitted parameters for the \ion{C}{1}
lines, and any others required to fit the data in the regions of those
lines, are given in Table \ref{tab:allparm}. $^{13}$\ion{C}{1} was
marginally detected in only the narrow component, N2.

The \ion{C}{1} Doppler parameter, $b_{\rm\, CI}=5.08\pm 0.24$ {\kms},
for component N1 corresponds, if the broadening is purely thermal, to a
temperature of over 16~000~K, so it is likely that bulk motions are
the major contributor. We cannot determine if it is a single component
with some bulk motion across or within it, or if the apparent width
arises because more than one component is present with separations
less than the instrument resolution. The populations of the three
\ion{C}{1} levels are not consistent with a single excitation
temperature - the best fit is $T_{\rm ex}=11.24\pm 0.34$~K, but the
observed \ion{C}{1}** column density is then too high relative to the
fit by $\Delta\log N=0.65$, i.e. over $4.5\sigma$.

The unresolved \ion{C}{1} in component N2 with $b_{\rm\, CI}=0.55\pm
0.13$ {\kms} has, for thermal line broadening, a kinetic temperature
$T=220$~K, with a $2\sigma$ upper limit $T<480$~K. This is much lower
than the 21\,cm spin temperature for the whole complex, and much
higher than the \ion{C}{1}*/\ion{C}{1} excitation temperature (which,
for this component, is $6.90^{+0.76}_{-0.62}$~K, in agreement with
\citet{Son94} to within the errors).

Component N3 is required only for the strongest two \ion{C}{1} lines
(at 1560 and 1656\,{\AA}). It is very broad, and all we can say from
its presence is that there are likely to be a number of weak
\ion{C}{1} components present over a velocity range $\sim
50$~km~s$^{-1}$ centered on the redshift given. They do not strongly
affect the derived parameters for the two narrower systems.

\section{The narrow CI component}
\label{narrowCI}

\subsection{Doppler width determination and its reliability}
\label{detdop}

The \ion{C}{1} lines in component N2 are unresolved, and their profiles
in the spectral data are close to the instrument profile, so we should
consider further whether or not these Doppler parameters and their
error estimates are reliable. If all the lines are unsaturated then
the Doppler parameters are not well constrained, as demonstrated by
\citet{Nar06}. However, it is possible to infer the Doppler
parameters for the unblended \ion{C}{1} lines to quite good precision
provided that not all of the lines are on the linear part of the
absorption line curve of growth, as was demonstrated in a similar
context by \citet{Jor09}. \citet{Strom48} describes the basic
technique as applied to line doublets, and this generalizes to
multiple lines, and multiple species. There are many other examples of
its use in this way, e.g. \citet{Mor75}, \citet{McC03} and
\citet{Cui05}.  For unresolved lines the minimum $\chi^2$ profile
fitting technique matches the equivalent widths of the lines, and so
if there are a number of saturated as well as unsaturated lines
present the Doppler parameter $b$ may be determined even where there
are possible blends with other components or ions from different
redshift systems.

\begin{figure}
\includegraphics[scale=0.6,angle=270]{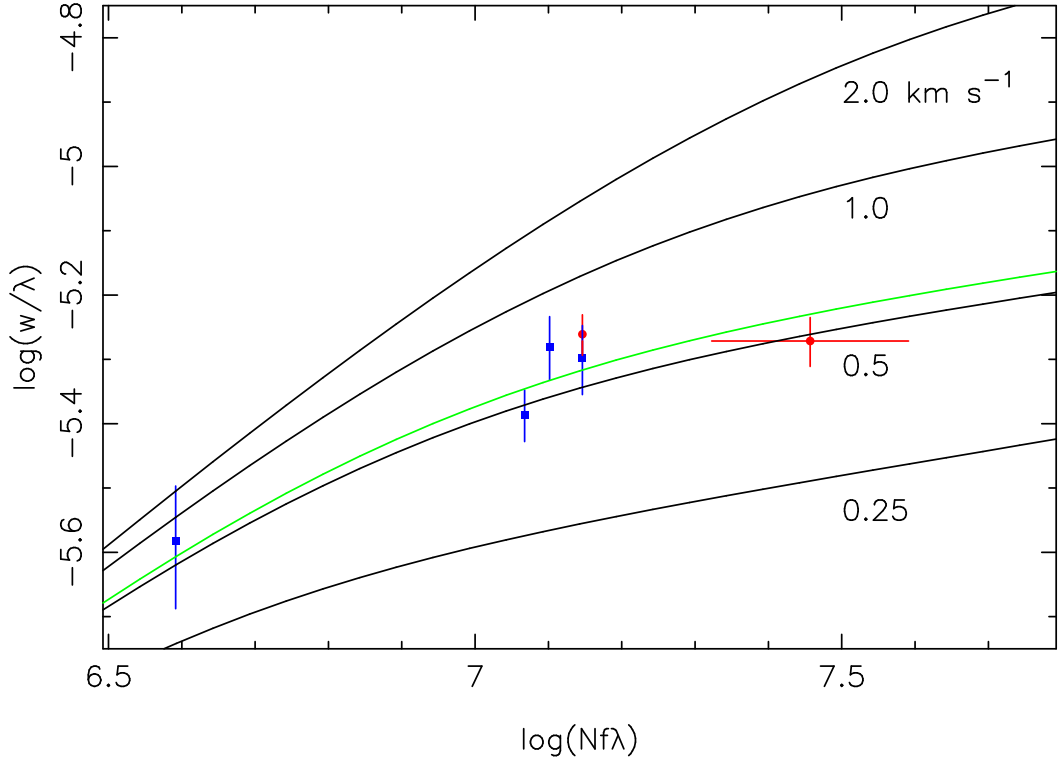}
\caption{Curves of growth for Doppler parameters $b=0.25$, 0.5, 1.0,
2.0 and 4.0~km~s$^{-1}$ (solid lines) and for the component B
best-fit value of $b=0.55$~km~s$^{-1}$ (green). Estimated ranges
for the equivalent widths $w$ of the \ion{C}{1} transitions at rest
wavelengths $\lambda=1656$, 1560, 1277, 1328 and 1280 give the
$w/\lambda$ values ordered from right to left. For \ion{C}{1} 1560 
both the HIRES (red filled circles) and UVES (blue squares)
equivalent 
width ranges are shown. The best estimate
column density $\log N=13.06$ was used to set the
$x$-position, and the error range correponding to the uncertainty in 
$\log N$ is shown against the 1656A line.\label{fig:cog}}
\end{figure}

To illustrate how it applies for the narrow \ion{C}{1} component
discussed here, {\reva and to highlight the circumstances under which 
we may derive reliable Doppler parameters and column densities from unresolved 
absorption lines,\finreva} we have generated a new continuum which contains all
absorption lines fitted apart from those of component N2 ground state
\ion{C}{1}. We then determined equivalent widths, with approximate
error estimates, for the component N2 \ion{C}{1} 1656, 1560, 1328, 1280
and 1277\,{\AA} transitions, and show these in Fig \ref{fig:cog} as
error ranges on a curve of growth where the \ion{C}{1} column density
is the best fit value from Table \ref{tab:allparm}.

{\reva It is clear from Fig. \ref{fig:cog} that the measured transitions with 
highest and the lowest oscillator strengths, i.e. 1656 and 1280, are
vital for constraining the Doppler parameter. 
The free parameter along the $x$-axis is the \ion{C}{1}
column density, and had the strongest line, \ion{C}{1}\,1656 not been 
measurable (e.g. through blending) then the remaining points adequately fit
the curves for any larger Doppler parameter - the \ion{C}{1} column density
could then be a factor of two lower, and the points on the linear part of the
curve of growth, and so the Doppler parameter constrained only by the
much larger instrument resolution value. 

If the weakest line used, \ion{C}{1}\,1280, had not been measured, then
there are much greater uncertainties in the column density, which could then be
well over a factor of ten higher with the consequence that the Doppler 
parameter would be somewhat lower. This would reduce the temperature upper
limit, and make any relative abundance analysis very uncertain.

These may be obvious points to make, but when using automated profile
fitting procedures one can forget about their inadequacies, or ignore
the large error estimates. It is important to check that the constraints
one obtains are based on the data, and not e.g. some approximation made
in estimating the errors in the parameters.\finreva}

The VPFIT
program provides error estimates from the diagonal terms in the
covariance matrix, and these are generally reliable for systems in
which the lines are resolved. For systems where the line widths are
considerably less than that of the instrument profile the reliability
of the estimates for the parameters and their errors has been
little investigated. We have done so here by generating 84 simulated
spectra with the same S/N as the original data using the line
parameters given in Table \ref{tab:allparm}. For each of these Voigt
profiles were fitted using the same fitting regions as in the original
data, and the resultant parameters compared with those used to
generate the spectra.

From these trials the distribution of both the log column
density and Doppler parameters alone shows that there is no
significant difference between the input and mean output values. For
the column density the mean and standard deviation are $\overline{\log
N(\ion{C}{1})}=13.077$ and $s=0.104$, compared with an input value of
${\log N(\ion{C}{1})}=13.064$ and VPFIT error estimate of $0.134$.
For the Doppler parameter input $b=0.553$, with VPFIT error $0.125$,
the mean estimate from the 84 trials is $\bar b=0.561$, with a
standard deviation of $0.082$. Therefore the difference between mean
values from the simulated spectra differ from the input values by
about the estimated error for those means (for the log column
densities the difference is 0.013 and error estimate 0.011; for the
Doppler parameters the corresponding quantities are 0.009 and 0.008).

From the same trials we find that the carbon isotope ratio and its
error estimate for this component
$\log\left(N(^{12}\ion{C}{1})/N(^{13}\ion{C}{1})\right)=1.28\pm 0.31$
are reliable. A $2\sigma$ lower limit for the $^{12}$C/$^{13}$C ratio is
5.

Further checks were undertaken to verify that there are no additional
uncertainties in the analysis of the data:

\begin{itemize}

\item The stopping criterion for the iterations involved in the
  fitting process was chosen to be when a change in $\chi^2<1.0\times
  10^{-5}$. For any search where convergence to the solution is slow
  this could result in e.g. the Doppler parameter being biased towards
  the initial guess relative to the true value. Tests showed this
  concern to be unfounded for the stopping criterion adopted.

\item Tests were run to verify that the adopted instrument resolution
  did not significantly affect the results. It is difficult to see how
  the resolution in each case could be worse than the slit-limited
  values which were adopted for the fits described above, but in
  conditions of good seeing the values of the FWHM appropriate for the
  spectra could be lower. We attempted to estimate the instrumental
  resolution by minimizing $\chi^2$ for profile fits to the
  \ion{Fe}{2} 2344, 2374, 2382, 2586 and 2600~A lines at $z=1.328$,
  and found somewhat broad minima with FWHM$\sim 5.8$~{\kms} for both
  the HIRES and UVES data. Adopting this value yields $\log
  N(\ion{C}{1})=13.07\pm 0.18$ with $b=0.47\pm 0.18$ for system N2.

\item \citet{Jen01} have estimated \ion{C}{1} oscillator strengths
  from interstellar medium absorption in early-type stars using HST
  STIS data.  Jenkins (private communication) has also suggested that
  the \ion{C}{1}~1560 oscillator strength, which is given by
  \citet{Mor03} as $f_{ik}=0.0774$, could be as high as
  $f_{ik}=0.1316$.  Voigt profile fits with the \citet{Jen01}
  $f$-values give results which differ by only a little from those
  obtained using the \citet{Mor03} values, with $b=0.53\pm
  0.13$~{\kms} and $\log N(\ion{C}{1})=12.96 \pm 0.17$, though the
  $\chi^2$ statistic for the fit is now somewhat too high.

\end{itemize}

From all these trials we conclude that the parameter estimates for
system N2 from VPFIT, for this mixture of saturated and unsaturated
lines, are reliable, though the VPFIT error estimates are too high by
a factor of $\sim 1.3$ - 1.5. In particular, the best estimate for the
\ion{C}{1} Doppler parameter from the trials is $b_{\rm\, CI}=0.553
\pm 0.082$. This corresponds to a temperature of $T=220$~K if the
broadening is thermal, and in any case the $2\sigma$ upper limit to
the kinetic temperature for this component is 480~K. Consequently 
the gas in component N2 is very likely to be a CNM.

\subsection{Physical conditions derived from \ion{C}{1}}

The closeness of the \ion{C}{1} excitation temperature
$6.90^{+0.76}_{-0.62}$~K to the microwave background temperature at
the system redshift, $T_{\rm CMB} = 2.725(1 + z_{\rm abs}) = 7.57$\,K,
allows us to place an approximate upper limit to the density in the
region, assuming that its temperature is significantly higher, since
the density must be low enough that collisional processes are
unimportant. We find a hydrogen number density $n_{\rm H}\simlt
3$\,cm$^{-3}$.
Under these circumstances, if most of the elements are neutral, we can
put a lower limit on the size of the cloud by estimating the amount of
neutral hydrogen associated with the cloud as follows: Assuming solar
relative abundances, [C/H]$_\odot = -3.61$, and therefore, $\log
n({\rm H})_{\rm cloud} = \log n({\rm CI})_{\rm cloud} + 3.61$.
Summing over the ground and excited states gives the total $\log
n({\rm CI})_{cloud} = 13.10$ cm$^{-2}$, therefore $N({\rm
  HI})_{cloud} = 10^{16.71}$cm$^{-2}$.  The cloud must therefore be
greater than $\ell = N({\rm H I})/n({\rm H I}) \simgt 1.7 \times
10^{16}$\, cm = 0.006 pc. 
This lower limit is sufficiently small that we would, without evidence
from H$_2$, be concerned that the \ion{C}{1} absorber only partially
covers the background source. We believe this is not the case, for the
reason given at the end of the next section (\ref{secH2}).

\section{Molecular hydrogen}
\label{secH2}

\subsection{Results from a single component fit}

Further evidence for CNM gas in DLA1331$+$170 stems from the detection
of {\Hmol} absorption lines at redshift $z=1.776553$ in the Lyman and
Werner bands with rest wavelengths from the Lyman limit to $\sim$ 1120
{\AA} {\citep{Cui05}}. These authors found the rotational level
populations between $J=0$ and $J=5$ to be consistent with a Boltzmann
distribution characterized by a single excitation temperature, 
$T_{\rm ex}$=152$\pm$10 K.  This differs from the Galaxy ISM where low
values of {\Tex} apply for $J$=0 to $J$=2, or $J=1$ to $J=3$, while
higher values of {\Tex} apply if higher values of $J$ are considered.
The standard interpretation of the ISM results is that for ortho- (odd $J$)
and para- (even $J$) H$_2$ separately the rotational
states with the lower values of $J$ are collisionally populated which
drives {\Tex} toward the kinetic temperature, $T$, while rotational
states with higher values of $J$ are populated both by UV pumping and
by the formation processes on grain surfaces which leave H$_{2}$ in
the $J$=4 state (Spitzer \& Zweibel 1974). The radiative processes are
dominant at high $J$ values because radiative self-shielding is less
important.  As a result, the constant value of {\Tex} in the case of
DLA1331$+$170 implies that {\Tex}=$T$ and that collisional processes
dominate radiative processes for \emph{all} of the rotational J levels
$J$=0 to $J$=5.

If collisional processes dominate for $J=4$ and 5, then we can place a
lower limit on the gas density. It must then exceed the critical
density $n_{\rm crit}$$\equiv$$A_{J,J-2}/q_{J,J-2}$, where $A_{J,J-2}$
and $q_{J,J-2}$ are the rates of de-excitation due to spontaneous
photon decay and collisions through transitions between the $J$ and
$J-2$ states. For the $J$=4$\rightarrow$2 transition, $n_{\rm crit}$
must exceed $\sim 5\times 10^{3}$ cm$^{-3}$ when $T$=150 K. The gas
giving rise to {\Hmol} absorption has some fraction of the total {\nh}
column density associated with it, so for this region {\nh} $\le
1.5\times$10$^{21}$ cm$^{-2}$. The corresponding length-scale of a
cloud with this volume density is $d\le 0.1$~pc, so much less than the
size of the optical continuum source of the background quasar, which
probably exceeds 0.5 pc. Therefore, unless the gas is in a thin sheet
perpendicular to the sightline, the smaller {\Hmol} absorbing cloud
cannot cover the continuum source, which it must do in order to
produce the saturated and damped absorption lines that arise in
DLA1331$+$170.

\begin{figure}
\includegraphics[scale=0.60,angle=90]{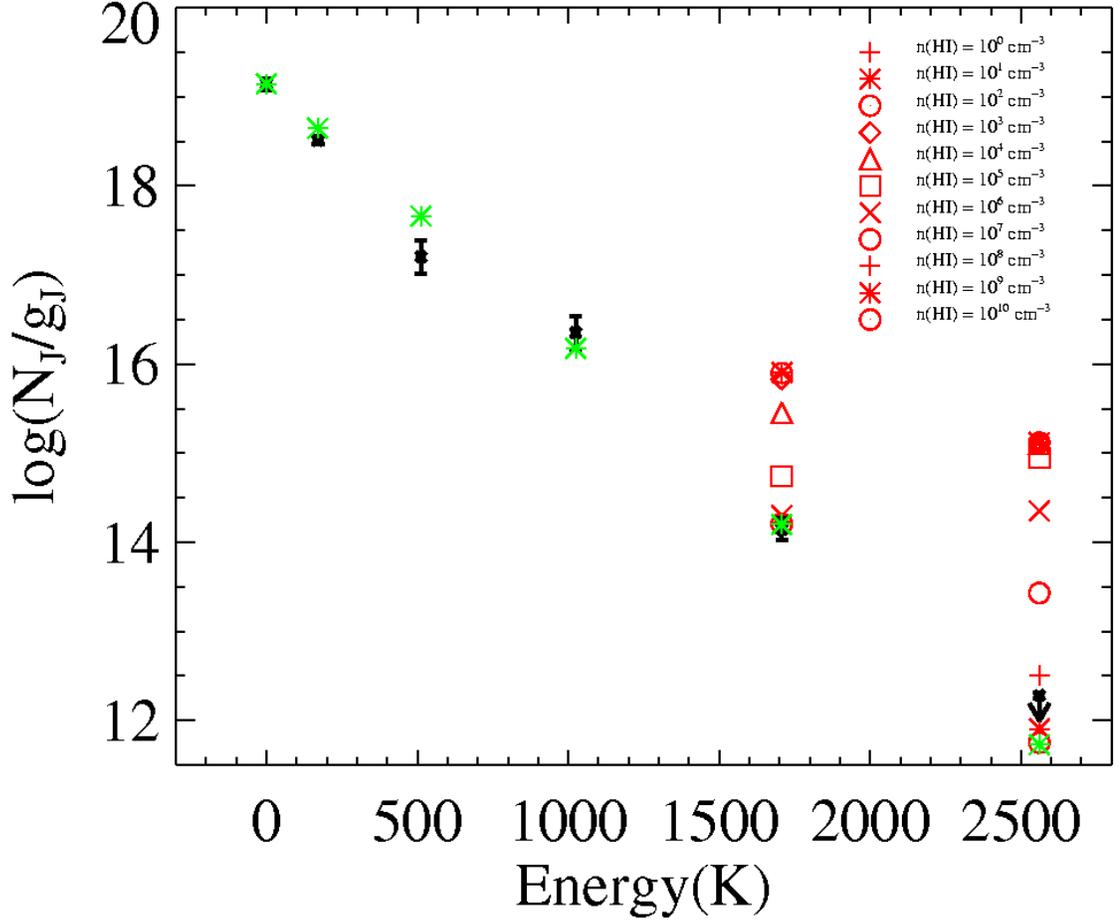}
\caption{The distribution of H$_2$ over $J$-levels 0 - 5. The abscissa
is the energy of the level relative to $J=0$, and the ordinate is the
column density divided by the level degeneracy. The black points with
error bars show the \citet{Cui05} values with error bars, and the
green points are the expected values for an excitation temperature of
$T_{\rm ex}=150$~K. The red points show the expected distribution for
a mixture of collisional excitation at this temperature and the
Haardt-Madau background radiation at redshift $z=1.7765$ for a range
of densities as indicated, with the highest points corresponding to 
the lowest densities.
\label{fig:level4}}
\end{figure}

For these reasons {\citet{Cui05}} assume that radiative rather than
collisional processes govern the population of the rotational
states up to $J\ge 4$.  But in that case the similarity between the
{\Tex} computed for states with $J\ge 4$ and $J<4$ (where {\Tex}=$T$)
would be a coincidence. To obtain the required level populations,
{\citet{Cui05}} find that at $\lambda$ $\approx$ 1000 {\AA}, $J_{\nu}$
= 6.7{$\times$}10$^{-23}$ {\junit}, which they claim is a reasonable
value for FUV background radiation at $z$ = 1.77, since if one assumes
a frequency dependence of ${\nu}^{-0.5}$, $J_{\nu}$ is in accord with
the value of 7.6{$\times$}10$^{-23}$ {\junit} at the Lyman limit
inferred from the proximity effect in the {\lya} forest. However, the
problem with this argument is that a power-law extrapolation across
the Lyman limit frequency, while valid for the quasar contribution to
the FUV background, ignores the dominant contribution of Lyman Break
galaxies at FUV frequencies.  According to {\citet{Haa01}},
$J_{\nu}$$\approx$3{$\times$}10$^{-20}$ {\junit} at $\lambda$
$\approx$ 1000 {\AA}, which is a factor of 500 higher than the
{\citet{Cui05}} estimate.  The effect of the {\citet{Haa01}}
background on the level populations is shown in Fig. \ref{fig:level4},
which plots the quantity $N_{J}/g_{J}$ versus $(E_{J}-E_{0})/k$, where
$N_{J}$ and $g_{J}$ are the column densities and degeneracies of the
$J^{th}$ rotational state, and $E_{J}$ is the corresponding energy
eigenvalue. 
We show the odd and even $J$ lower level populations at 
the same excitation temperature since that is what the data indicates, 
but have not considered processes linking the two.

As a result our analysis of the H$_{2}$ absorption lines arising in
DLA1331$+$170 has raised the following dilemma: If the $J$=0
{$\rightarrow$} 5 rotational states are populated by collisions, then
the scale length of the {\Hmol} absorbing gas would be too small.  On
the other hand if the rotational states with $J$ $\ge$ 4 are
radiatively populated by plausible FUV radiation fields, the predicted
values of $N_{J}/g_{J}$ would considerably exceed the values inferred
from a single {\Hmol} cloud model fit.

\subsection{Interpreting the H$_2$ data: Multiple component molecular hydrogen}

The presence of velocity components N1, N2 and N3 hints at a 
possible solution to
this dilemma; namely, the bulk of the H$_{2}$ gas resides in the
narrow-lined component N2, with the broader components N1 and N3
containing a smaller amount. Consider the evidence. First, our best
estimate for the kinetic temperature of component N2, $T\sim 200$~K
(from \ion{C}{1} if the line broadening is thermal), is consistent
with the H$_2$ excitation temperature, {\Tex} = $152 \pm 10$~K
obtained by \citet{Cui05}.  Second, the redshift difference between
component N2 and {\Hmol} absorption, 3.0 {\kms}, is less than the
difference between {\Hmol} and either of the other two \ion{C}{1}
components. Although the Doppler parameter for the H$_{2}$ absorption
complex, $b=13.9 \pm 0.5$~km~s$^{-1}$ is considerably greater than
that of component N2, we suggest that while component N2 is embedded in
the more turbulent medium that dominates the velocity structures of
the H$_{2}$ absorption lines with higher $J$ values, it none-the-less
contains the bulk of the H$_{2}$ gas. Consequently, while component N2
is responsible for most of the damping-wing absorption produced for
the $J$=0 and 1 lines, its low Doppler velocities imply that the
absorption from states with higher $J$ values will be dominated by
absorption arising from the more turbulent velocity structures N1 and
N3. Stated differently, the values of $N_{J}$ for $J$ $\ge$ 3 may have
been under-estimated by the failure to identify the low equivalent
widths generated by component N2. As a result, the $N_{J}/g_{J}$ versus
$(E_{J}-E_{0})/k$ curve may in fact be consistent with FUV radiation
intensities due to the {\citet{Haa01}} backgrounds and possibly low
rates of local star formation. Finally, it is well-known that because
\ion{C}{1} and H$_2$ are photoionized and photo-dissociated by photons
of similar energy, they are likely to be co-spatial and hence show 
similar velocity structure.

\begin{figure}
\epsscale{.66}
\plotone{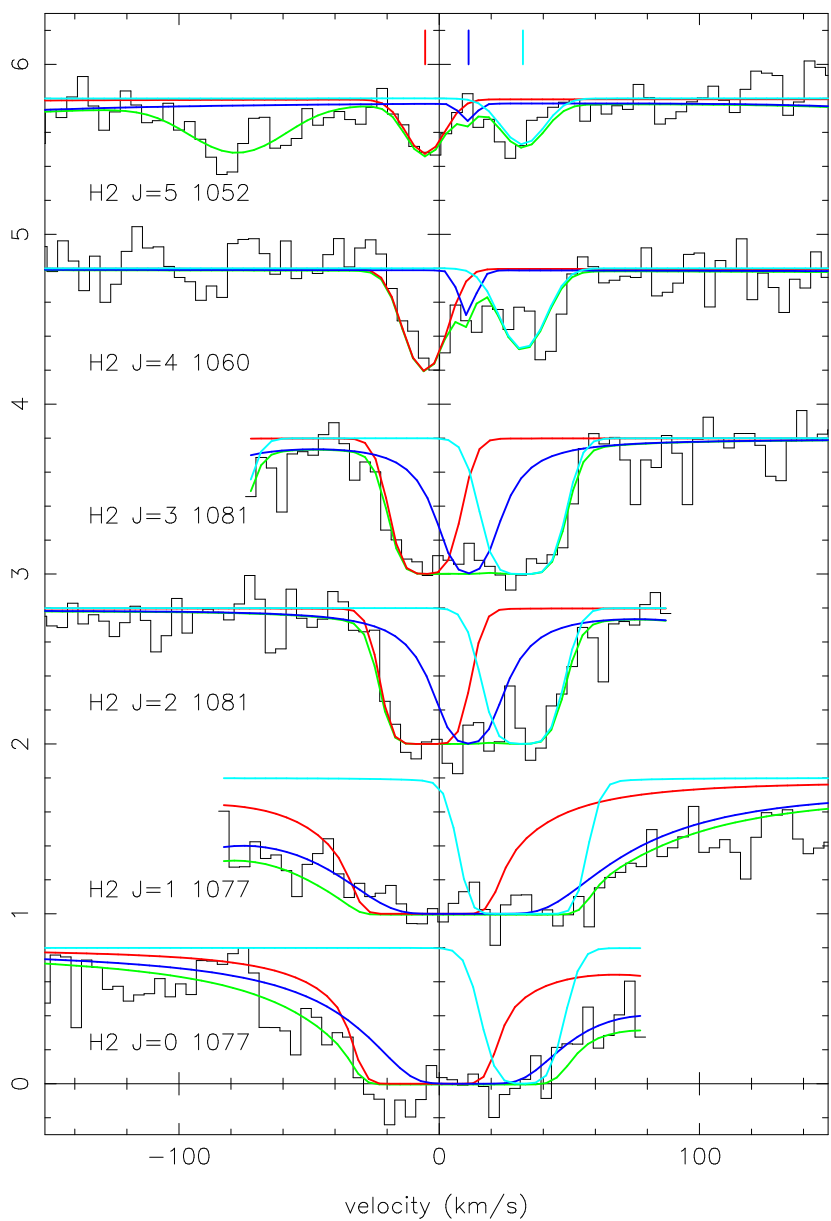}
\caption{\footnotesize Representative H$_2$ absorption lines from 
the $J=0$ to 5
levels for the three-component model described in the text. The data
are shown in black, the overall fit in green, and the individual
contributions from the $z=1.7763702$, 1.7765246 and 1.7767176
components in red, blue and turquoise respectively. The zero-velocity 
reference is $z=1.776420$, the continuum is set to 0.8 everywhere, and 
successive lines are offset in $y$ by 1.0. 
The narrow component (blue) dominates the $J=0$ and 1 lines,
but is only a minor contributor for $J=4$ and 5. \label{fig:H2}}
\end{figure}

To investigate this possibility further we have performed our own
analysis of the \citet{Cui05} STIS data, assuming a component
structure suggested by the \ion{C}{1} results. We have taken three
components for H$_2$, two at the redshifts corresponding to the
\ion{C}{1} lines in components N1 and N2. Since the redshift of the
third \ion{C}{1} component, N3, is not well constrained the third H$_2$
component redshift {\reva (denoted N3$^\prime$, $z=1.7767176\pm 0.0000067$) 
\finreva} was 
determined by the profile fit to the molecular
hydrogen lines.  The Doppler parameters were constrained so they are
physically consistent with those of the corresponding \ion{C}{1}
i.e. $b_{\rm\, CI}\leq b_{\rm\, H_2}\leq \sqrt{6}\, b_{\rm\, CI}$, with
the limits corresponding to purely turbulent and purely thermal broadening. For
the narrow component the adopted Doppler parameter was $b_{\rm\,
  H_2}=1.0$~\kms. The parameters for the profile fits obtained with
these constraints are given in Table \ref{tab:H2parm}, which also
lists the transitions used in the fits. In this case the low $J$
absorption is dominated by component N1, while the higher $J$ is
dominated more by the other two velocity components. This is
illustrated in Fig. \ref{fig:H2} using selected transitions for 
each $J$ level.

Our three-component fit to the H$_2$ lines is by no means unique, but
it does serve to provide a more physically consistent picture of the
complex, at least within the rather large errors. The single-component
fit giving an apparent single excitation temperature for all
excitation levels arises simply because different components dominate 
different $J$-values, and is unlikely to reflect the true physical 
conditions in the individual components.

The molecular hydrogen results also suggest that here, as in the case
of \citet{Jor09}, the narrow components inferred for \ion{C}{1} are
unlikely to be an artefact caused by a small covering factor for the
absorbing gas, since the corresponding saturated H$_2$ have zero
residual intensities.

\section{Ionized heavy elements}

\subsection{Singly ionized species}
\label{secWNM}

\begin{figure}
\epsscale{.66}
\plotone{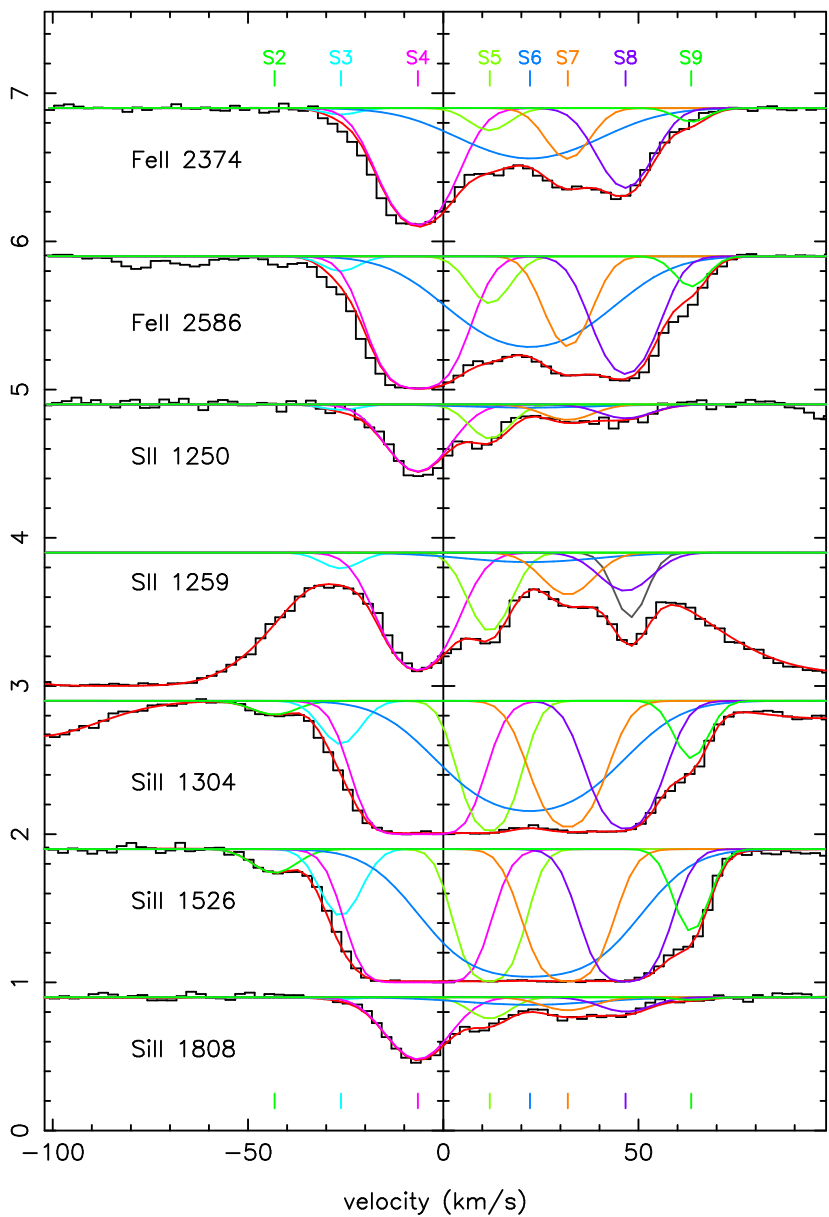}
\caption{\footnotesize The singly ionized complex centered on redshift
$z=1.776420$ showing the total fits in red. These include unrelated blends 
and the component structure for the lines in the complex.
Component velocities and 
labels (see Table \ref{tab:WNMz}) are marked in the corresponding colours.
The continuum level is set at 0.9 and each line biassed up by an integer amount
to separate them.
\ion{Fe}{2}\,2586 was not included in the fitted regions because of the
possibility that the weak structure seen in the blue wing may extend into 
the line itself, but is shown here as an example of the consistency of the
fit obtained using other lines.
The dark grey component shown 
in the \ion{S}{2}\,1259 profile is \ion{Si}{2}\,1260 at $z=1.7748722$.
\label{fig:IIstrong}}
\end{figure}

\begin{figure}
\epsscale{.66}
\plotone{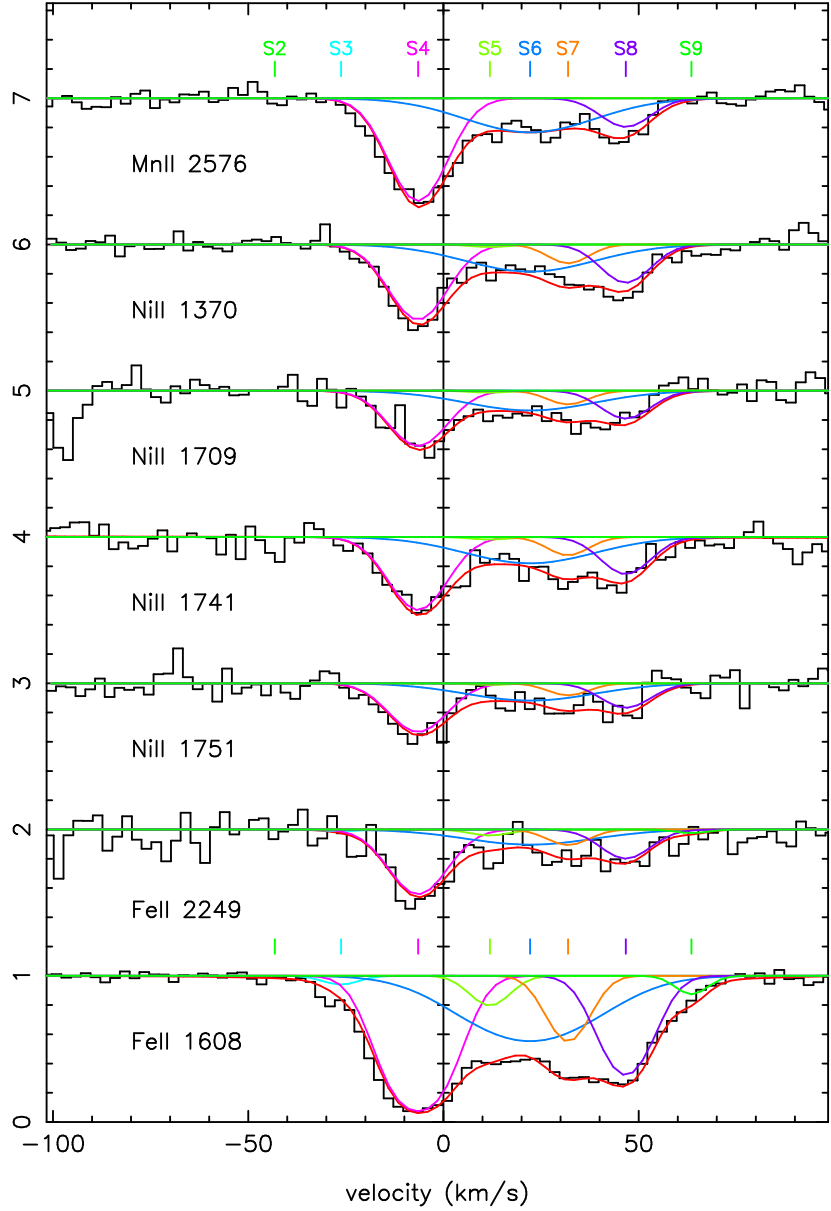}
\caption{\footnotesize The singly ionized complex centered on redshift
$z=1.776420$ showing the total fits in red, 
and the component structure for the lines in the complex for the weaker lines.
Component velocities and 
labels (see Table \ref{tab:WNMz}) are marked in the corresponding colours.
The continuum and zero level for the \ion{Fe}{2}\,1608 are unity and zero, 
as shown, but for the other lines the vertical scale has been stretched 
by a factor of four, so the zero levels are four units below the continuum
which has been biassed upward in each case by an integer amount to separate the lines.
\label{fig:IIweak}}
\end{figure}

{\reva While DLA$1331+170$ presents clear evidence for the presence of
cold gas, it is difficult to determine how much of the neutral hydrogen is associated
with it, and so what the overall cold gas fraction of the DLA may be.
In an attempt to clarify the overall structure of the DLA
we analyze the non-neutral heavy elements.

Some of the \ion{H}{1} will be associated with singly ionized
heavy elements e.g. \ion{C}{2},
\ion{Mg}{2}, \ion{Si}{2}, \ion{Fe}{2} etc., so that
at least part of the component structure of \ion{H}{1} will be traced 
by these ions. For
the $z=1.77642$ complex discussed here, the dominant components of all the
available lines of \ion{C}{2} and \ion{Mg}{2} are either
strongly saturated or too weak to be measurable, so they provide
little useful information. For this complex
\ion{Si}{2}~1808, \ion{S}{2}~1250, 1259, \ion{Mn}{2}~2576,
\ion{Fe}{2}~1608, 2249, 2374 and \ion{Ni}{2}~1370, 1709, 1741, 1751
have the right combination of oscillator strength and column density
to be useful, and the saturated lines \ion{Si}{2}~1260, 1304, 1526
were also used to provide some additional constraints. Some other
potentially useful lines were found to be blended with strong lines
from other redshift systems, so were omitted from the analysis. All the 
elements used, with the exception of sulphur, have ionization potentials
for I-II in the range 7.4 - 8.2 eV, and II - III from 15.6 - 18.2 eV,
so they should arise predominantly in the same regions. For sulphur the 
corresponding potentials are 10.4 and 23.3 eV, so there could be some 
differences but we failed to find any. For \ion{Al}{2} there is only
a single transition available, at~1670A, and it was so saturated that
the column densities of the dominant components could not be determined.

Since we are interested in physical entities, we have fitted the
multiple ions to common redshifts, and constrained the Doppler
parameters assuming that bulk motions have a Gaussian distribution,
and so the bulk and thermal motions for each ion add in
quadrature. Then, for each component, $b=\sqrt{b_{\rm turb}^2 +
2kT/m}$, where $b_{\rm turb}$ is the turbulent (bulk) component, $k$
Boltzmann's constant, $m$ the ion mass and $T$ the temperature. In
terms of the VPFIT program used, this involves fitting $b_{\rm turb}$
and $T$ for at least two ions of different mass at the same redshift
simultaneously, with the constraints that both variables
are non-negative. The error estimates from the program then apply to
$b_{\rm turb}$ and $T$, not to the Doppler parameters for the individual 
ions.

From the fitting analysis 9 separate velocity components are needed,
labelled S1 - S9 in increasing redshift order. Details of these are
given in Table \ref{tab:WNMz},  along with parameters for other systems 
with lines in
the fitting regions used. All the lines fitted were available in the UVES 
spectrum, and for the HIRES data FeII\,2249, SiII\,1808, 
NiII\,1751, NiII\, 1709,
FeII\,1608 and part of SiII\,1526 were included as well. 
The fitted line profiles and component structure 
are shown against the UVES data in Figs \ref{fig:IIstrong} and 
\ref{fig:IIweak}. An additional component at $z=1.7748722$ was
included, since \ion{Si}{2}~1260 from that system is blended with
\ion{S}{2}~1259 from the main complex.  Its structure was determined
using \ion{C}{2}~1334, \ion{Mg}{2}~2796, 2803, \ion{Si}{2}~1193, 1304,
1526. \ion{Fe}{2} is not seen in this component, with an upper limit
$\log N(\ion{Fe}{2})<11.5$ (cm$^{-2}$, $2\sigma$). 

The component structure given in Table \ref{tab:WNMz} is somewhat different
from that given by \citet{Pro97}, who illustrate a fitted profile to the 
SiII\,1808 line. A large part of the difference probably comes about 
because we have used the additional UVES spectrum and fitted
many transitions where they used just one.

For determining
which ions provide the strongest constraints it can be useful to have
Doppler parameter error estimates for individual ions rather than
those for the variables $b_{\rm turb}$ and $T$. Indicative values are
given Table \ref{tab:WNMz}. Here the error estimates given for the individual
Doppler parameters for each ion were computed by assuming that the
best-fit $b$-value is the tied one, but computing the errors as if the
tied constraint were not applied. In some cases the error exceeded the
Doppler parameter, so there is no useful constraint. These are
indicated by a dash in the table.

As can be seen from Table \ref{tab:WNMz}, for the component structure
inferred here, bulk motions dominate in several cases. These are S3, S4, S5, S6,
S8 and S9, where the Doppler parameters have similar values for all
ions. For S1 and S7, however, thermal broadening is more important,
though the temperature estimate for S7, at 95~000~K, is
somewhat higher than we might expect. There are several components
across the velocity range, and so, partly because most of the systems
are now blended and partly because the mass range over which reliable
Doppler parameters can be obtained is at most from $^{28}$Si to
$^{56}$Fe i.e. a factor of less than 2, the error estimates for both
$b_{\rm turb}$ and $T$ are rather large. For component S7, the $1\sigma$
error estimate is almost 70~000~K, so within the errors the temperature
could be only a few $\times 10^4$~K.

Most temperature errors are smaller, and in two cases (S4 \& S5) the
$2\sigma$ upper limit indicates temperatures of $\sim 2\times 10^4$~K
or less.  In general the temperatures are at least consistent with the
notion that the material is in the WNM with temperatures of $\sim 1$ -
$3\times 10^4$~K. However, the errors are large, and so CNM gas is an 
equally viable interpretation. 
\finreva}

\subsection{More highly ionized heavy elements}
\label{secdbion}

{\reva
We may explore the structure of more highly ionized species, though
because they have a range of ionization potentials, we cannot necessarily 
treat them as coming predominantly from the same 
physical regions. The motivation for doing this is to check whether or not
comparison of the column densities of close ions, such as \ion{Al}{2} and
\ion{Al}{3}, provides a reliable indicator of the ionization state
of the gas. 

The doubly ionized species detected in our observed range are 
\ion{Si}{3}\,1206 and \ion{Al}{3}\,1854, 1862. \ion{Si}{3} has only a single 
line which is saturated,
in the Ly$\alpha$ forest and close to the strong damped Ly$\alpha$ associated 
with the system, so it provides little useful information. \ion{Fe}{3}\,1122
is potentially observable, but in a region in the Ly$\alpha$ forest ($<3120$A)
where the S/N in the data is poor. Consequently we
have used only the \ion{Al}{3} doublet to investigate the component structure 
here. The results are given in Table \ref{tab:AlIII}, and the component 
structure shown against other ions in Fig. \ref{fig:ccpts}.

Similarly we can investigate the structure of triply ionized species, 
such as \ion{Si}{4} and \ion{C}{4} through their doublet absorption lines. 
In this case we choose \ion{Si}{4} only, since its ionization potential 
(45.1eV) is less than that of \ion{He}{2} (54.4eV), whereas that of \ion{C}{4} 
(64.5eV) is above the \ion{He}{2} value, and therefore the two species may not
 be predominantly in the same regions. The \citet{Pro99} and 
\citet{Wol00} studies also show that while there is a general similarity
in the \ion{Si}{4} and \ion{C}{4} velocity structure, the correspondence
is not perfect. The results for the Voigt profile fits to the 
\ion{Si}{4} 1393, 1402 doublet 
are given in Table \ref{tab:AlIII}.
\finreva}

\section{Comparing the various components}

\subsection{Comparing profiles}
\begin{figure}
\epsscale{.66}
\plotone{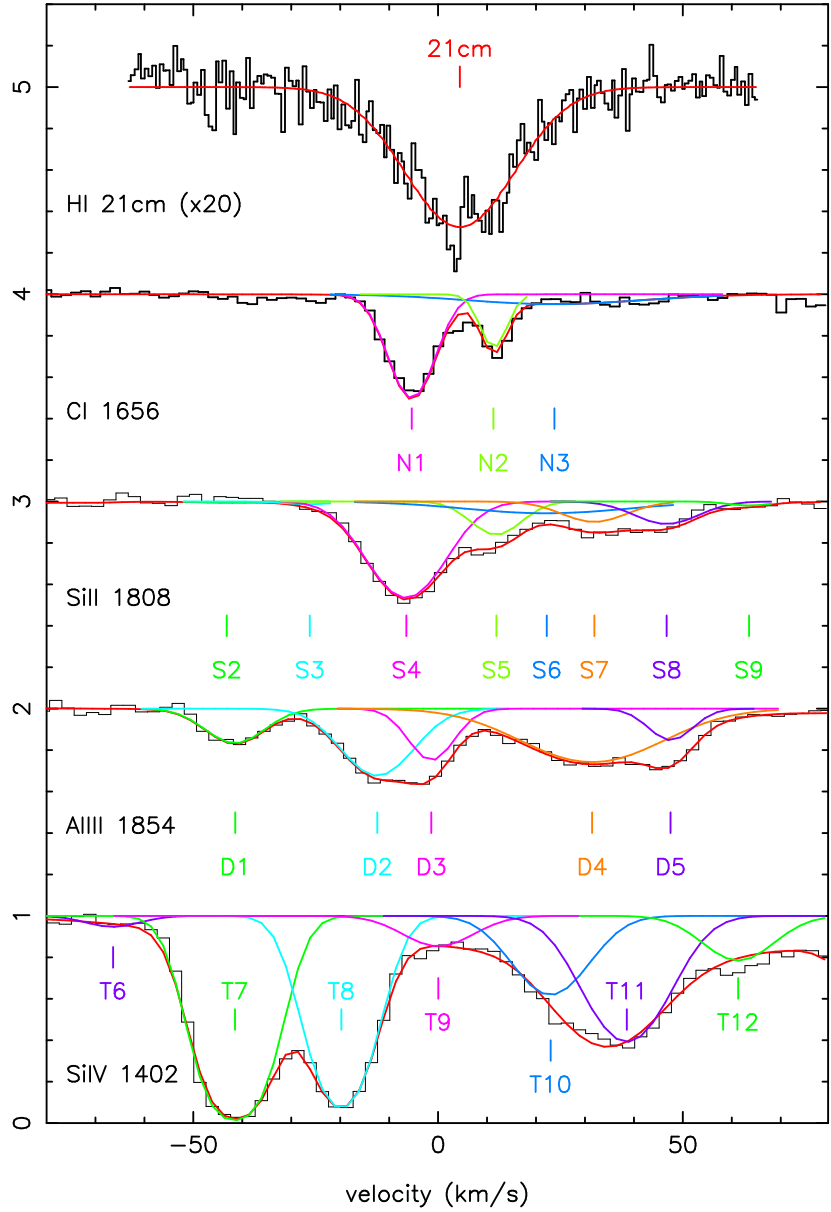}
\caption{\footnotesize The velcity structure of the \ion{H}{1} 21cm, \ion{C}{1}\,1656, \ion{Si}{2}\,1808 \ion{Al}{3}\,1854 and \ion{Si}{4}\,1402 absorption lines on a common velocity scale. The total Voigt profile fits are shown in red, and individual components for the various ions shown in colours corresponding to the labels given in Tables \ref{tab:allparm}, \ref{tab:WNMz} and \ref{tab:AlIII}.  
\label{fig:ccpts}}
\end{figure}

{\reva
Figure \ref{fig:ccpts} shows the \ion{H}{1}~21cm profile with
representative lines from neutral, singly and more highly ionizied heavy 
elements. This is quite instructive, since there are some features which may 
be common to more than one ionization level and others which seem to 
correspond less well.

The \ion{H}{1}~21cm profile stands out as having little in common with 
any of the others. While we have no direct measure of what the velocity 
distribution of the damped Ly$\alpha$ absorption is from the UVES
spectrum, if the heavy element abundances are not too disparate between
velocity components it would be surprising if it did not roughly follow
that of the the higher column density heavy element neutral 
and singly ionized species. Therefore we expect it to follow S4 - S8 
reasonably well, with some contribution from N1 and N2. 

The \ion{H}{1}~21cm line is  broad
and, at the S/N available, smooth, and is consistent with the
absorption arising in a single cloud at a temperature of $T=15 000$\,K.
The difference in shape from that expected from the heavy elements, and 
the redshift of the maximum absorption
from any of the components, make it difficult to escape the conclusion
that it and the Ly$\alpha$ absorption have little to do with each other.
This reinforces
the notion, raised in the context of other objects by \citet{Cur07} and 
\citet{Kan09}, that the 21cm and Ly$\alpha$ absorption lines 
do not arise in the same region(s), possibly
because the radio source is extended and so it and the optical source 
are not coincident. Consequently
the derivation of a spin temperature by using the radio 21\,cm and the
optical Ly$\alpha$ lines may well be inappropriate.

The various ionization stages of the heavy element lines show considerable
overlap, especially for the lower ionization states. However, from Voigt 
profile fitting each ionization stage independently,
it is not always clear for adjacent ionization levels what
velocity structure they have in  common. This is
most readily seen from Fig. \ref{fig:ccpts}, with details of redshift
and Doppler parameters in Tables \ref{tab:allparm}, \ref{tab:WNMz} and 
\ref{tab:AlIII}. Comparative details extracted from these tables are given, 
in redshift order, in Table \ref{tab:vcpt}. In this table we have chosen 
\ion{Si}{2}  as the representative of the singly ionized components, 
since the silicon and aluminium
atomic masses are close in value and so the Doppler parameters of any cospatial
components should be nearly the same. 

In Table \ref{tab:vcpt} possible goupings of systems are indicated as those
between horizontal lines, and for those where the redshifts and Doppler
parameters are compatible right hand brackets. Others within groupings
may have significant components in common, especially where the Doppler
parameters are significantly different. And, of course, the component 
structure inferred by assuming the minimum
number of components consistent with the data may not reflect reality, and in any case does 
not necessarily lead to unique results (see Kirkman \& Tytler, 1997, for a discussion applied to the Ly$\alpha$ forest). However it is not 
immediately 
clear what assumptions we should make to attempt to disentangle individual
physical components. 

One possibility is to assume the absorption for each component comes
from a single region or cloud. If material with different 
ionization levels is cospatial, then the redshifts should be the same and the
Doppler parameters should reflect the same bulk flows and temperatures.
A variant of this was adopted by \citet{Mil10}, who assumed that 
the same redshifts and Doppler parameters applied for neutral to 
doubly-ionized species in a $z_{\rm abs}\sim 2$ sub-DLA. While this
approach might be appropriate for high $b$-values, we doubt its validity
for narrow lines where thermal broadening may be important since a
$\sqrt{T/m}$ term which depends on the atomic mass $m$ will make a 
significant contribution. Even allowing for this, there may well be 
temperature structure linked with 
ionization structure within a cloud which receives ionizing radiation from 
an external source. This may provide an explanation for the significant 
difference in Doppler parameters between components N2 and S5, which share 
a common redshift so could well be parts of the same structure. If there is 
any \ion{Al}{3} associated with this component, it can not be separated from
the much broader D4.

Similarly, it is also tempting to suggest that N1 and S4 are parts of the
same cloud, though the Doppler parameters are rather large and difficult
to reconcile with a turbulent+thermal model. Also, here the component 
velocities relative to the reference redshift, $\Delta v$, are significantly different. However, the velocity 
difference ($1.1\pm 0.2$ {\kms}) is small, and consistent with that which 
might be expected from a photdissociation region \citep{Hol99}, implying that the 
physical association picture may still be correct. So in this case a thermal
and bulk flow separation within the entity could provide an explanation. 
Once again, any associated \ion{Al}{3} is not discernible, but could be hidden in components D2 and D3.

Despite the above considerations, we further explored possible constraints 
by forcing different ionization levels to have the same 
redshifts, but allowing the Doppler parameters to differ between them.
In the end this is not useful because the blending of components leads to 
ill-constrained solutions, and the column densities in for the components
forced into the blends have large errors. For example, because of the small but significant difference in the redshift centroids between N1 and S4, the 
singly ionized component
S4 had to be divided into two, one corresponding to the \ion{C}{1} component 
N1 and another, S4$^\prime$, still close to the initial S4. The column density errors in the new part of S4  corresponding to N1 were high (generally $> 0.2$ dex
from VPFIT, so likely to be underestimates), and provided little new 
information. Similarly the errors in the \ion{Al}{3} column density in 
the component forced to the redshift of S4$^\prime$ were high, formally a 
factor of 1000, and so not useful. Some attempts on subsets where the Doppler
parameters were also constrained yielded similar results - the column density
errors were again too large for the values to be useful.

While it is clear that some \ion{Al}{3} is associated with the singly 
ionized components, some is also predominantly associated with \ion{Si}{4}.
A similar association of some of the \ion{Al}{3} with higher ionization 
components was noted 
in GB$1759+7539$ by \citet{Pro02}. A consequence of this is that associating 
the total \ion{Al}{3} column density with the \ion{Al}{2} (or its proxy, 
\ion{Si}{2}) will lead to an over-estimate of the amount of ionized material 
associated directly 
with the \ion{H}{1} regions. Unfortunately for DLA$1331+170$ 
we were unable to associate velocity components across ionization states
to obtain good relative column densities. Similar ambiguities arise in
a detailed study of a nearby multi-phase Galactic interstellar cloud 
\citep{Neh08} for similar reasons - there are too many indistinct 
components in velocity space.

\finreva}

\subsection{Cold gas}

{\reva \citet{Wol03b} have shown that the bulk of the medium showing
\ion{C}{2}$^*$ absorption must be cold in general. Since we have
another indicator of cold gas in \ion{C}{1} which is largely distinct
from that traced by the singly ionized heavy elements, we could, at
least in principle, check this directly. In DLA$1331+170$ the
\ion{C}{2}$^*$~1335 transition is blended with a Ly$\alpha$ forest
line at $z=2.0506465$, so the weaker \ion{C}{2}$^*$~1335 lines are
partly masked out. There are sharper \ion{C}{2}$^*$~1335 components within
the Ly$\alpha$ profile, and they are compatible with both the N1-N3
and S2-S9 structure. The \ion{C}{2}$^*$~1335/Ly$\alpha$ profile is shown in
\citet{Jor10}.

Fitting models based upon the velocity structure of the neutral and
singly ionized lines, along with the Ly$\alpha$ forest line, results in 
each case in summed
column densities for
\ion{C}{2}$^*$ which are smaller than that quoted in Table 1 in
\citet{Wol03a}.  They quote $\log N(\ion{C}{2}^*)=14.05\pm 0.05$,
while we find that, for the sum $\log N(\ion{C}{2}^*)=13.59\pm 0.09$
(neutral \ion{C}{1} model) or $13.57\pm 0.55$ (singly ionized
model). 
\finreva}

The relationship between the \ion{C}{1} and H$_2$ is also not
completely clear.  The H$_2$ excitation temperature of $152\pm 10$~K
given by \citep{Cui05} is consistent with the kinetic $2\sigma$
temperature limit of $< 480$~K we find for the \ion{C}{1} associated
with component B. However their estimate is based on a
single-component model, while the \ion{C}{1} which we expect should be
co-spatial, shows at least three velocity components.  We have shown
in section \ref{secH2} that a three-component fit for the H$_2$ is
compatible with the \ion{C}{1} and does, within the rather large
errors, allow the H$_2$ upper level populations to be consistent with
the FUV background radiation.
The temperature results are summarized in Table \ref{tab:alltemp}. The
para- and ortho- H$_2$ excitation temperatures agree at about the
combined $1\sigma$ level, though there may be a small radiative term
enhancing the $J=3$ level.

Because the isotope shift for $^{13}$C relative to $^{12}$C is
potentially measurable, $^{13}$C was included in in the analysis, though
it made little difference to the final results. Our
lower limit of $^{12}$C/$^{13}$C$> 5$ ($2\sigma$) for component N2 is
considerably smaller than the lower limit $^{12}$C/$^{13}$C$> 80$
given by \citet{Lev06} from their analysis of the \ion{C}{1} 1560 and
1656 ground state and fine structure transitions at redshift
$z=1.150789$ towards HE$0515-4414$.  We suspect that the errors they
give may be optimistic, since our reanalysis of the archive data for
that object indicates that $^{12}$C/$^{13}$C$> 22$, with the main
constraint coming from the \ion{C}{1}$^*$ fine structure
lines. However, we are still some way from being able to use carbon
isotope abundance limits to investigate chemical enrichment mechanisms
at high redshifts. \citet{Pra96} have shown that for timescales $<
5\times 10^9$ years this ratio is $>100$. For very low metallicity
stars, which may be more appropriate for the DLA$1331+170$ case, that
ratio can be even higher, $\simgt 1000$ \citep{Lau07}.

Given the upper limit on the temperature, and the upper
limit on the density of $n_{\rm H}\simlt 3$~cm$^{-3}$ we could
consider, as in \citet{Jor09}, whether or not the narrow component
cloud could be gravitationally confined. However, it is not clear how
much of the \ion{H}{1} is associated with the cold component here, so
we are not able to draw any firm conclusions. If we guess that, say,
10\% of the \ion{H}{1} seen is associated with the \ion{C}{1},
corresponding to $\log N({\rm HI})=20.17$ (cm$^{-2}$) then we find
that the cloud is unlikely to be gravitationally bound.

\section{Conclusions}

We have presented here a second DLA with a hidden, cold component
revealed through
narrow \ion{C}{1} lines. The first, described by \citet{Jor09}, used
curve-of-growth techniques on an isolated \ion{C}{1} component. Here
we have shown that, even if there is a small amount of blending, it is
possible to determine line widths, and temperature limits (in this
case $<480$\,K), well below spectral resolution if the spectra contain
a number of lines of the same ion with differing oscillator strengths,
provided that they span different parts of the classical curve of
growth. This is neither new nor surprising, but it is comforting to
know that what are usually thought of as 'profile fitting' packages
don't rely exclusively on the line profile information, and that the
Doppler parameter error estimates are reliable. 

We note that while the spin, excitation, and kinetic temperatures
within the absorbing system differ considerably those
temperatures that should be physically consistent indeed are, as in
the narrow component kinetic temperature and the lower $J$ molecular
hydrogen excitation.  Even given this agreement, the single velocity
component analysis of the higher $J$ {\Hmol} excitation yielded level
populations which are too low for the expected background
radiation levels. This conflict is removed by decomposing the H$_2$
into three velocity
components, corresponding to those found for \ion{C}{1}. This example
shows the potential importance of considering multi-component models to
help interpret observed molecular hydrogen level populations.

{\reva
Attempts to determine temperatures for singly ionized components using 
transitions from ions with a range of atomic masses were not generally
successful. For the strong components the transitions of the lightest
ions, such as \ion{C}{2}, are saturated, and so weaker lines from \ion{Si}{2}
and \ion{Fe}{2} have to be used, and even for these blending adds 
significantly to the uncertainties. The weak limits we have obtained
are consistent with temperatures of up to a few $\times 10^4$\,K.

While common components across various close ionization levels are clearly
evident, the detailed overall correspondence in velocity and Doppler parameter between
the various ionization levels I - IV is not very good. In particular,
there are differences between the close ions \ion{Si}{2} and \ion{Al}{3}, with
some of the \ion{Al}{3} associated instead with higher ionization material.
Consequently, attempts to derive
ionization corrections using ratios using these could well give 
misleading results unless there is a detailed
study of corresponding velocity structures. Associating all the 
\ion{Al}{3} with the lower ionization species is likely to overestimate the
amount of ionized gas directly associated with the \ion{H}{1} gas. 
While the results here apply to 
a DLA system, they suggest that the interpretation of abundances 
relative to hydrogen in sub-DLAs, where ionization corrections are 
more crucial, may need to be re-examined. This will not be very 
important for DLAs, where the average heavy element abundances in each 
system are reasonably well estimated by comparing singly ionized heavy 
elements with \ion{H}{1}, since
the total column densities are usually well constrained (see e.g. the 
VPFIT version 9.5 program documentation, section 11.4) and the ionization 
corrections small. 

The profile and component disparity between the heavy elements which
should trace the neutral hydrogen reasonably well, the neutral and singly 
ionized components, and the 21cm absorption profile, reinforces the
suggestion that the 21cm and Ly$\alpha$ lines are only weakly related.
This may be due to the fact that the background radio emission is more 
spatially extended than the optical source, and so the sightlines sampled 
are not the 
same. This leads to the conclusion that, if this is a typical example, 
the individual 21cm spin temperatures 
determined by comparing the 21cm and Ly$\alpha$ lines are physically 
irrelevant.
\finreva}

\acknowledgements

\noindent{\it Acknowledgements}

We are grateful to Xavier Prochaska for his very helpful comments on a
draft version of this paper, and Gary Ferland and Wim Ubachs for
clarifying aspects of molecular hydrogen astrophysics for us, and
a referee for encouraging us to perform an analysis of the
ionized heavy elements. RFC is
grateful to the Leverhulme Trust for the award of an Emeritus
Grant. RAJ acknowledges support from the UK Science \& Technology
Facilities Council. AMW acknowledges support by the NSF through grant
AST-0709235. MTM thanks the Australian Research Council for a QEII
Research Fellowship (DP0877998).

\clearpage

\clearpage

\clearpage

\clearpage


\clearpage

\clearpage

\clearpage

\clearpage

%

\clearpage

\begin{table}
\begin{center}
\caption{CI and blends\label{tab:allparm}}
\null
{\scriptsize\renewcommand{\arraystretch}{0.78}
\begin{tabular}{lccrrccl}
\tableline\tableline
Ion&$z$&$\pm$&\multicolumn{1}{c}{$\phantom{2}b$}&\multicolumn{1}{c}{$\pm$}&$\log N$&$\pm$&
Notes\\ 
\tableline
Fe~I   &0.7446128&0.0000006& 1.20& 0.18&12.04&0.02&Fe~I~2484 near C~I*~1560\\ 
Al~III &1.3253472&0.0000014& 6.26& 0.32&12.30&0.02&Al~III~1862 near C~I~1560\\ 
Al~III &1.3281613&0.0000072& 7.17& 0.86&12.12&0.06&\\ 
Al~III &1.3282625&0.0000048& 6.22& 0.78&12.21&0.05&\\ 
Al~III &1.3285315&0.0000014&11.87& 0.27&12.55&0.01&\\ 
Al~III &1.3288447&0.0000057& 9.65& 1.15&11.76&0.04&\\ \tableline 
C~I    &1.7763702&0.0000009& 5.08& 0.24&13.08&0.02&{\bf N1} ($-5.4\pm 0.1$~\kms)\\
C~I*   &1.7763702&         & 5.08&     &12.59&0.02&\\ 
C~I**  &1.7763702&         & 5.08&     &11.90&0.14&\\ 
C~I    &1.7765246&0.0000014& 0.55& 0.13&13.06&0.13&{\bf N2} ($11.3\pm 0.2$~\kms)\\
C~I*   &1.7765246&         & 0.55&     &12.05&0.07&\\ 
C~I**  &1.7765246&         & 0.55&     &\multicolumn{2}{r}{$<11.70$}&\\ 
$^{13}$C~I&1.7765246&      & 0.53&     &11.78&0.28&\\
C~I    &1.7766400&0.0000467&24.45& 6.18&12.51&0.11&{\bf N3} ($23.8\pm 5.1$~\kms)\\ 
C~I*   &1.7766400&         &24.45&     &\multicolumn{2}{r}{$<12.20$}&\\ 
C~I**  &1.7766400&         &24.45&     &\multicolumn{2}{r}{$<12.25$}&\\ \tableline
C~IV   &1.7858576&0.0000180&28.88& 2.38&12.93&0.03&C~IV~1550 near Al~III~1854\\ 
C~IV   &1.7861859&0.0000117&10.49& 2.06&12.43&0.13&\\ 
C~IV   &1.7864237&0.0000116&16.49& 1.75&12.80&0.04&\\ 
C~IV   &1.7869053&0.0000121&14.83& 1.75&12.62&0.05&\\ 
C~IV   &1.7871942&0.0000187&15.63& 2.70&12.47&0.07&\\ 
H~I    &1.9174518&0.0000518&16.90& 9.29&11.93&0.23&Ly$\alpha$ in C~I~1277 region\\ 
H~I    &1.9178188&0.0000201& 7.20& 3.74&11.86&0.16&\\ 
H~I    &1.9183670&0.0000841&10.65&11.33&11.86&0.43&\\ 
 ??    &1.9238783&0.0000317& 7.78& 2.01&12.89&0.25&Line in C~I~1280 region\\ 
 ??    &1.9239821&0.0000106& 6.60& 0.93&13.23&0.11&\\ 
H~I    &1.9242860&0.0000042&11.48& 0.71&13.32&0.03&\\ 
 ??    &1.9243053&0.0000064& 1.21& 3.54&12.86&2.02&\\ 
 ??    &1.9244645&0.0000067& 2.21& 2.12&12.25&0.09&\\ 
H~I    &1.9247054&0.0000138& 9.95& 2.44&12.18&0.08&\\ 
C~IV   &1.9661609&0.0000093&16.53& 1.57&12.47&0.04&C~IV~1550 near C~I~1656\\ 
H~I    &2.0359392&0.0000731&37.31&12.66&12.14&0.17&Ly$\alpha$ in C~I~1328 region\\ 
H~I    &2.0369588&0.0000458&19.80& 7.66&12.03&0.24&\\ 
H~I    &2.0376443&0.0000388&55.83& 9.36&12.91&0.09&\\ 
\tableline
\end{tabular}
}
\tablecomments{Doppler parameters $b$ are in {\kms}, and column densities are
$\log N$ cm$^{-2}$. Error estimates are $1\sigma$, and upper limits are
$2\sigma$. The \ion{C}{1} wavelengths used were those for $^{12}$C except where
the isotope is given explicitly. Ly$\alpha$ lines with $b<10$~km~s$^{-1}$ are probably
unidentified heavy elements, and are marked '??'. Their parameters
were determined assuming that the rest wavelength and oscillator
strength are the same as for Ly$\alpha$. For the C~IV and Al~III
doublets both lines were included in the fits, and for Fe~I at redshift $z=0.774613$
the lines
at 2167, 2484 and 2719 were used. Component labels are shown in bold, and the velocities given for the three identified components are relative to a reference redshift $z=1.77642$.}
\end{center}
\end{table}

\clearpage

\begin{table}
\begin{center}
\caption{Multicomponent molecular hydrogen column densities\label{tab:H2parm}}
\null
{
\begin{tabular}{cccccccl}
\tableline\tableline
Component&\multicolumn{2}{c}{N1}&\multicolumn{2}{c}{N2}&\multicolumn{2}{c}{N3$^\prime$}\\
$z$&\multicolumn{2}{c}{1.7763702}&\multicolumn{2}{c}{1.7765246}&\multicolumn{2}{c}{1.7767176}&Lines used in fit\\
$b$(H$_2$)  (km\,s$^{-1}$)&8.7&$\pm 0.5$&1.0&-&10.0&$\pm 0.7$&\\
\tableline
$\log N(J$=0)&18.56&$\pm 0.28$&18.99&$\pm 0.16$&15.69&$\pm 0.63$&{\footnotesize 1049 1077 1092}\\
$\log N(J$=1)&18.90&$\pm 0.22$&19.46&$\pm 0.08$&17.39&$\pm 0.40$&{\footnotesize 932 1049 1051 1064 1077}\\
&&&&&&&{\footnotesize 1078 1092}\\
$\log N(J$=2)&16.68&$\pm 0.24$&18.44&$\pm 0.11$&16.10&$\pm 0.13$&{\footnotesize 927 987 1005 1016 1051}\\
&&&&&&&{\footnotesize 1053 1064 1066 1079 1081}\\
&&&&&&&{\footnotesize 1096}\\
$\log N(J$=3)&15.94&$\pm 0.14$&18.20&$\pm 0.14$&16.07&$\pm 0.14$&{\footnotesize 934 935 942 960  987 995}\\
&&&&&&&{\footnotesize 997 1006 1017 1041 1053}\\
&&&&&&&{\footnotesize 1056 1081 1096 1099}\\
$\log N(J$=4)&14.95&$\pm 0.06$&14.42&$\pm 0.35$&14.81&$\pm 0.06$&{\footnotesize 1017 1035 1047 1060 1074}\\
&&&&&&&{\footnotesize 1085 1099}\\
$\log N(J$=5)&14.46&$\pm 0.07$&13.61&$\pm 0.33$&14.42&$\pm 0.07$&{\footnotesize 996 997 1006 1017 1048}\\
&&&&&&&{\footnotesize 1052 1065 1079 1089}\\
$T_{\rm ex}(J$=0\,\&\,2) (K)&86&\rlap{\raisebox{0.2ex}{$^{+14}$}}\raisebox{-0.2ex}{{$_{-10}$}}&
177&\rlap{\raisebox{0.2ex}{$^{+30}$}}\raisebox{-0.2ex}{{$_{-22}$}}&$\simgt 200$&\\
\tableline
\end{tabular}
}
\tablecomments{Column densities are log cm$^{-2}$. The error estimates are $1\sigma$, and for the (log) column densities are likely to be underestimates if they exceed $\sim 0.3$. The excitation temperature for component N3$^{\prime}$ is effectively unconstrained - a $1\sigma$ lower limit is given}
\end{center}
\end{table}

\clearpage

\begin{table}
\begin{center}
\caption{Singly ionized species and blends\label{tab:WNMz}}
\null
{\footnotesize\renewcommand{\arraystretch}{0.65}
\begin{tabular}{lccrrcclcccc}
\tableline\tableline
Ion&$z$&$\pm$&\multicolumn{1}{c}{$\phantom{2}b$}&\multicolumn{1}{c}{$\pm$}&$\log N$&$\pm$&Sys&\multicolumn{3}{l}{$\Delta v$ ({\kms})}\\
&&&&&&&$b_{\rm turb}$&$\pm$&$T$ ($10^4$K)&$\pm$\\ 
\tableline
C II  &1.7748722&0.0000014& 5.36& (0.42)&13.20&0.04&{\bf S1}&\multicolumn{3}{l}{$-167.1$}\\
MgII  &1.7748722&         & 3.76& (0.28)&12.11&0.01&  0.0&  0.9&2.1&1.1\\
SiII  &1.7748722&         & 3.50& (0.64)&12.27&0.03&\\ \cline{1-1}\cline{8-11}
SiII  &1.7760202&0.0000039& 6.46& 0.52&12.70&0.03&{\bf S2}&\multicolumn{3}{l}{$-43.2\pm 0.4$}\\ \cline{1-1}\cline{8-11}
SiII  &1.7761772&0.0000078& 5.33& (0.89)&13.21&0.10&{\bf S3}&\multicolumn{3}{l}{$-26.2\pm 0.9$}\\
S II  &1.7761772&         & 5.20& (3.50)&13.44&0.13&  4.3&  4.2&1.7&5.8\\
FeII  &1.7761772&         & 4.82& (2.23)&12.46&0.14&\\ \cline{1-1}\cline{8-11}
SiII  &1.7763599&0.0000013& 9.53& (0.24)&15.07&0.01&{\bf S4}&\multicolumn{3}{l}{\phantom{0}$-6.5\pm 0.2$}\\
S II  &1.7763599&         & 9.53& (0.48)&14.86&0.01&  9.5&  0.4&0.0&1.3\\
MnII  &1.7763599&         & 9.53& (1.08)&12.16&0.03&\\
FeII  &1.7763599&         & 9.53& (0.26)&14.30&0.02&\\
NiII  &1.7763599&         & 9.53& (0.73)&13.10&0.03&\\ \cline{1-1}\cline{8-11}
SiII  &1.7765303&0.0000028& 5.67& (0.99)&14.35&0.09&{\bf S5}&\multicolumn{3}{l}{\phantom{$-$}$11.9\pm 0.3$}\\
S II  &1.7765303&         & 5.67& (1.02)&14.32&0.10&  5.7&  4.1&0.0&0.6\\
FeII  &1.7765303&         & 5.67& (3.80)&13.07&0.27&\\
NiII  &1.7765303&         & 5.67&  -  &11.43&1.01&\\ \cline{1-1}\cline{8-11}
SiII  &1.7766255&0.0000297&22.78& (2.34)&14.38&0.18&{\bf S6}&\multicolumn{3}{l}{\phantom{$-$}$22.2\pm 3.2$}\\
S II  &1.7766255&         &22.77&  -  &13.73&0.79& 22.6&  4.7&1.1&39\\
MnII  &1.7766255&         &22.72&(15.29)&12.00&0.06&\\
FeII  &1.7766255&         &22.71& (6.51)&13.99&0.08&\\
NiII  &1.7766255&         &22.71& (9.38)&12.98&0.08&\\ \cline{1-1}\cline{8-11}
SiII  &1.7767156&0.0000043& 8.11& (1.53)&14.23&0.12&{\bf S7}&\multicolumn{3}{l}{\phantom{$-$}$31.9\pm 0.5$}\\ 
S II  &1.7767156&         & 7.66& (2.04)&14.03&0.16&  3.1&  4.2&9.5&6.6\\
FeII  &1.7767156&         & 6.14& (1.46)&13.52&0.11&\\
NiII  &1.7767156&         & 6.03& (3.70)&12.33&0.15&\\ \cline{1-1}\cline{8-11}
SiII  &1.7768523&0.0000029& 8.00& (0.55)&14.27&0.03&{\bf S8}&\multicolumn{3}{l}{\phantom{$-$}$46.7\pm 0.3$}\\
S II  &1.7768523&         & 7.97& (2.53)&13.99&0.07&  7.7&  0.8&0.8&2.0\\
MnII  &1.7768523&         & 7.86& (6.00)&11.51&0.13&\\
FeII  &1.7768523&         & 7.86& (0.67)&13.87&0.04&\\
NiII  &1.7768523&         & 7.85& (1.82)&12.72&0.05&\\ \cline{1-1}\cline{8-11}
SiII  &1.7770083&0.0000030& 4.01& (0.42)&13.32&0.04&{\bf S9}&\multicolumn{3}{l}{\phantom{$-$}$63.5\pm 0.3$}\\
FeII  &1.7770083&         & 3.85& (1.48)&12.76&0.09&  3.7&  2.0&0.4&2.4\\
\tableline
H I   &1.8757384&0.0000243&23.39& 0.94&14.33&0.09&\\
H I   &1.8775792&0.0000028&35.60& 0.48&14.03&0.01&\\
H I   &1.8760396&0.0001075&17.46& 6.86&13.25&0.53&\\
H I   &1.9779551&0.0000061&21.84& 0.88&12.97&0.01&\\
H I   &1.9799652&0.0000161&26.69& 2.63&12.70&0.03&\\
??    &2.6737241&0.0000526&49.43& 5.17&12.73&0.06&\\
\tableline
\end{tabular}
}
\tablecomments{\footnotesize
Component labels are given in bold. Error estimates are $1\sigma$, and
those in parentheses are indicative Doppler parameter errors if the
ion were treated without reference to others in the system (see text
for details). Other details as for Table \ref{tab:allparm}.}
\end{center}
\end{table}

\clearpage

\begin{table}
\begin{center}
\caption{AlIII \& SiIV absorption\label{tab:AlIII}}
\null
{\scriptsize\renewcommand{\arraystretch}{0.78}
\begin{tabular}{lccrrcccl}
\tableline\tableline
Ion&$z$&$\pm$&\multicolumn{1}{c}{$\phantom{2}b$}&\multicolumn{1}{c}{$\pm$}&$\log N$&$\pm$&Sys&$\Delta v$ ({\kms})\\ 
\tableline
AlIII&1.7760362&0.0000047& 7.12& 0.95&11.99&0.04&{\bf D1}&$-41.5\pm 0.5$\\ 
AlIII&1.7763051&0.0000172& 8.93& 1.56&12.40&0.11&{\bf D2}&$-12.4\pm 1.9$\\
AlIII&1.7764071&0.0000098& 4.74& 1.69&12.08&0.21&{\bf D3}&\phantom{0}$-1.4\pm 1.1$\\
AlIII&1.7767114&0.0000107&18.85& 1.41&12.57&0.04&{\bf D4}&\phantom{$-$}$31.5\pm 1.2$\\
AlIII&1.7768598&0.0000052& 4.56& 1.67&11.83&0.13&{\bf D5}&\phantom{$-$}$47.5\pm 0.6$\\
\tableline
SiIV&1.7748504&0.0000012& 6.62& 0.16&13.36&0.01&{\bf T1}&$-169.53$\\ 
SiIV&1.7750657&0.0000223&10.35& 3.50&12.81&0.36&{\bf T2}&$-146.27$\\
SiIV&1.7750956&0.0000062& 1.30& 1.03&12.39&0.14&{\bf T3}&$-143.04$\\
SiIV&1.7752779&0.0001684&23.70&10.30&12.77&0.48&{\bf T4}&$-123.34$\\
SiIV&1.7752931&0.0000043& 5.12& 1.13&12.71&0.12&{\bf T5}&$-121.71$\\
SiIV&1.7758053&0.0000093& 6.25& 1.92&11.89&0.08&{\bf T6}&$-66.4\pm 1.0$\\
SiIV&1.7760357&0.0000012& 6.82& 0.22&13.90&0.03&{\bf T7}&$-41.5\pm 0.1$\\
SiIV&1.7762370&0.0000014& 6.52& 0.30&13.63&0.02&{\bf T8}&$-19.8\pm 0.2$\\
SiIV&1.7764205&0.0000098& 8.94& 3.32&12.47&0.14&{\bf T9}&\phantom{$-$0}$0.1\pm 1.1$\\
SiIV&1.7766329&0.0000282& 9.42& 2.71&12.97&0.25&{\bf T10}&\phantom{$-$}$23.0\pm 3.0$\\
SiIV&1.7767771&0.0000166&10.16& 1.95&13.29&0.12&{\bf T11}&\phantom{$-$}$38.6\pm 1.8$\\
SiIV&1.7769879&0.0000085& 8.97& 2.69&12.66&0.31&{\bf T12}&\phantom{$-$}$61.3\pm 0.9$\\
SiIV&1.7773069&0.0000041&10.41& 1.65&13.29&0.21&{\bf T13}&\phantom{$-$}$95.8$\\
SiIV&1.7773179&0.0000567&29.11&50.81&12.96&0.18&{\bf T14}&\phantom{$-$}$96.9$\\
SiIV&1.7774448&0.0000038& 2.17& 1.05&12.59&0.08&{\bf T15}&$110.6$\\
SiIV&1.7776712&0.0000063& 5.84& 1.77&12.52&0.20&{\bf T16}&$135.1$\\
SiIV&1.7778277&0.0000541&15.81& 5.97&12.40&0.33&{\bf T17}&$152.0$\\
\tableline 
\end{tabular}
}
\tablecomments{Error estimates are $1\sigma$. The velocities $\Delta v$ are relative to a reference redshift $z=1.77642$. See Table \ref{tab:allparm} for
further details.}
\end{center}
\end{table}

\clearpage

\begin{table}
\begin{center}
\caption{Velocity components compared\label{tab:vcpt}}
\null
{\footnotesize\renewcommand{\arraystretch}{0.65}
\begin{tabular}{lccrrccllc}
\tableline\tableline
Ion&$z$&$\pm$&\multicolumn{1}{c}{$\phantom{2}b$}&\multicolumn{1}{c}{$\pm$}&$\log N$&$\pm$&Sys&$\Delta v$ ({\kms})\\
\tableline
SiIV  &1.7758053&0.0000093& 6.25& 1.92&11.89&0.08&{\bf T6}&$-66.4\pm 1.0$\\ \cline{1-1}\cline{8-9}
SiII  &1.7760202&0.0000039& 6.46& 0.52&12.70&0.03&{\bf S2}&$-43.2\pm 0.4$\\
AlIII &1.7760362&0.0000047& 7.12& 0.95&11.99&0.04&{\bf D1}&$-41.5\pm 0.5$&\lower 0.25ex\hbox{$\rceil$}\\
SiIV  &1.7760357&0.0000012& 6.82& 0.22&13.90&0.03&{\bf T7}&$-41.5\pm 0.1$&\raise 0.25ex\hbox{$\rfloor$}\\ \cline{1-1}\cline{8-9}
SiII  &1.7761772&0.0000078& 5.33& 0.89&13.21&0.10&{\bf S3}&$-26.2\pm 0.9$\\ \cline{1-1}\cline{8-9}
SiIV  &1.7762370&0.0000014& 6.52& 0.30&13.63&0.02&{\bf T8}&$-19.8\pm 0.2$\\ \cline{1-1}\cline{8-9}
AlIII &1.7763051&0.0000172& 8.93& 1.56&12.40&0.11&{\bf D2}&$-12.4\pm 1.9$\\ \cline{1-1}\cline{8-9}
SiII  &1.7763599&0.0000013& 9.53& 0.24&15.07&0.01&{\bf S4}&\phantom{0}$-6.5\pm 0.2$\\
C~I   &1.7763702&0.0000009& 5.08& 0.24&13.08&0.02&{\bf N1}&\phantom{0}$-5.4\pm 0.1$\\ \cline{1-1}\cline{8-9}
AlIII &1.7764071&0.0000098& 4.74& 1.69&12.08&0.21&{\bf D3}&\phantom{0}$-1.4\pm 1.1$&\lower 0.25ex\hbox{$\rceil$}\\
SiIV  &1.7764205&0.0000098& 8.94& 3.32&12.47&0.14&{\bf T9}&\phantom{$-$0}$0.1\pm 1.1$&\raise 0.25ex\hbox{$\rfloor$}\\ \cline{1-1}\cline{8-9}
C~I   &1.7765246&0.0000014& 0.55& 0.13&13.06&0.13&{\bf N2}&\phantom{$-$}$11.3\pm 0.2$\\
SiII  &1.7765303&0.0000028& 5.67& 0.99&14.35&0.09&{\bf S5}&\phantom{$-$}$11.9\pm 0.3$\\ \cline{1-1}\cline{8-9}
SiII  &1.7766255&0.0000297&22.78& 2.34&14.38&0.18&{\bf S6}&\phantom{$-$}$22.2\pm 3.2$&\lower 0.25ex\hbox{$\rceil$}\\ 
SiIV  &1.7766329&0.0000282& 9.42& 2.71&12.97&0.25&{\bf T10}&\phantom{$-$}$23.0\pm 3.0$\\
C~I   &1.7766400&0.0000467&24.45& 6.18&12.51&0.11&{\bf N3}&\phantom{$-$}$23.8\pm 5.1$&\raise 0.25ex\hbox{$\rfloor$}\\ \cline{1-1}\cline{8-9}
AlIII &1.7767114&0.0000107&18.85& 1.41&12.57&0.04&{\bf D4}&\phantom{$-$}$31.5\pm 1.2$\\
SiII  &1.7767156&0.0000043& 8.11& 1.53&14.23&0.12&{\bf S7}&\phantom{$-$}$31.9\pm 0.5$\\ \cline{1-1}\cline{8-9}
SiIV&1.7767771&0.0000166&10.16& 1.95&13.29&0.12&{\bf T11}&\phantom{$-$}$38.6\pm 1.8$\\ \cline{1-1}\cline{8-9}
SiII  &1.7768523&0.0000029& 8.00& 0.55&14.27&0.03&{\bf S8}&\phantom{$-$}$46.7\pm 0.3$&\lower 0.25ex\hbox{$\rceil$}\\
AlIII &1.7768598&0.0000052& 4.56& 1.67&11.83&0.13&{\bf D5}&\phantom{$-$}$47.5\pm 0.6$&\raise 0.25ex\hbox{$\rfloor$}\\ \cline{1-1}\cline{8-9}
SiIV  &1.7769879&0.0000085& 8.97& 2.69&12.66&0.31&{\bf T12}&\phantom{$-$}$61.3\pm 0.9$\\
SiII  &1.7770083&0.0000030& 4.01& 0.42&13.32&0.04&{\bf S9}&\phantom{$-$}$63.5\pm 0.3$\\
\tableline
\end{tabular}
}
\tablecomments{\footnotesize
Component labels are given in bold. Systems which may be associated are grouped
between horizontal lines, and those for which the redshifts and Doppler parameters are compatible
are shown linked with a right bracket. See Table \ref{tab:allparm} for other details.}
\end{center}
\end{table}

\clearpage

\begin{table}
\begin{center}
\caption{Temperature estimates\label{tab:alltemp}}
\null
{\renewcommand{\arraystretch}{1.0}
\begin{tabular}{lccrrccl}
\tableline\tableline
Ion \& method&$z$&$\pm$&\multicolumn{1}{c}{$\phantom{2}b$}&\multicolumn{1}{c}{$\pm$}&$T$&$\pm$&
Notes\\ 
\tableline
H$_2$ T$_{{\rm ex}\  J=0\,\&\,2}$&1.7765246&0.0000014&1.00&  - &177&30&Component N2\\ 
H$_2$ T$_{{\rm ex}\  J=1\,\&\,3}$&1.7765246&0.0000014&1.00&  - &225&28&Component N2\\ 
C{\small I} T$_{\rm ex}$    &1.7763702&0.0000009&5.08&0.24&10.4&0.5&Component N1\\
C{\small I} T$_{\rm ex}$    &1.7765246&0.0000014&0.55&0.13&\phantom{1}7.2&0.8&Component N2\\
C{\small I} Doppler         &1.7765246&0.0000014&0.55&0.08&220&$<480$&Component N2\\
\tableline
\end{tabular}
}
\tablecomments{Error estimates are $1\sigma$, and upper limits are
$2\sigma$.}
\end{center}
\end{table}

\end{document}